\documentclass[12pt]{article}

\usepackage{epsf}
\usepackage{amssymb} % \mathfrak, \mathbb
\usepackage{graphicx}
\makeatletter
\@addtoreset{equation}{section}

\renewcommand{\thefootnote}{\fnsymbol{footnote}}
\makeatother

\begin{document}
\thispagestyle{empty}

%%%%%%%% margin %%%%%%%%%%%%%%%%%%%%%%%%%%%
%\def\baselinestretch{1.2}
%\parskip 6 pt
%\marginparwidth 0pt
%\oddsidemargin  0pt
%\evensidemargin  0pt
%\marginparsep 0pt
%\topmargin   -0.5in
%\textwidth   6.5in
%\textheight  9.0 in

%%%%%%    personal macro %%%%%%%%%%%%%%%%%%%
\newcommand{\nn}{\nonumber}

\begin{flushright}
TIT/HEP--495 \\
{\tt hep-th/0306077} \\
June, 2003 \\
\end{flushright}
\vspace{3mm}

\begin{center}
{\Large
{\bf Domain Wall Junction 
in ${\cal N}=2$ 
Supersymmetric 
} 
\\
\vspace{2mm}{\bf 
QED in four dimensions 
}
} 
\\[12mm]
\vspace{5mm}

\normalsize
  {\large \bf 
  Kazuya~Kakimoto}
\footnote{\it  e-mail address: 
kakimoto@th.phys.titech.ac.jp
}  
~and~~  {\large \bf 
Norisuke~Sakai}
\footnote{\it  e-mail address: 
nsakai@th.phys.titech.ac.jp
} 

\vskip 1.5em

{ \it Department of Physics, Tokyo Institute of 
Technology \\
Tokyo 152-8551, JAPAN  
 }
\vspace{15mm}
{\bf Abstract}\\[5mm]
{\parbox{13cm}{\hspace{5mm}
%%%%%%%%%%%%%%%%%%%%%%%%%%%%%%%%%%%%%%%%%%%%%%%%%%
%%%%%%%%%%
An exact solution of domain wall junction is obtained in 
${\cal N}=2$ supersymmetric (SUSY) QED with three massive 
hypermultiplets. 
The junction preserves two out of eight SUSY. 
Both a (magnetic) Fayet-Iliopoulos (FI) term and complex masses for 
hypermultiplets are needed to obtain the junction 
solution. 
There are zero modes corresponding to spontaneously broken 
translation, SUSY, and $U(1)$. 
All broken and unbroken 
SUSY charges are explicitly worked out in the Wess-Zumino 
gauge in ${\cal N}=1$ superfields as well as in components. 
The relation to models in five dimensions is also clarified. 
%%%%%%%%%%%%%%%%%%%%%%%%%%%%%%%%%%%%%%%%%%%%%%%%%%
}}
\end{center}
\vfill
\newpage
\setcounter{page}{1}
\setcounter{footnote}{0}
\renewcommand{\thefootnote}{\arabic{footnote}}

%%%%%%%%%%%%%%%%%%%%%%%%%%%%%%%%%%%%%%
\section{Introduction}\label{INTRO}
%%%%%%%%%%%%%%%%%%%%%%%%%%%%%%%%%%%%%%
%\vspace{5mm}

In recent years,  models with extra dimensions have 
attracted much attention 
\cite{LED}, \cite{RS}. 
In this  brane-world scenario, our world is assumed 
to be realized on extended topological defects 
such as domain walls or junctions. 
On the other hand, supersymmetry (SUSY) provides 
the most promising idea to build realistic unified 
theories beyond the standard model \cite{DGSWR}. 
Brane-world scenario in the supersymmetric theories 
can provide an opportunity for a realistic model 
building on walls and/or junctions. 
Moreover, it can offer a possible explanation 
of SUSY breaking \cite{RSsb}--\cite{KT}, 
in particular 
by means of the coexistence of walls 
\cite{MSSS}, \cite{MSSS2}. 
SUSY has been useful to obtain solutions of walls 
and junctions as BPS states, which 
preserve a part of SUSY \cite{WittenOlive}. 

Domain walls can conserve half of the SUSY, and are called 
${1 \over 2}$ BPS states. 
They have been extensively studied in globally supersymmetric theories 
\cite{CQR}, \cite{DW}. 
More recently,  
an exact BPS wall solution in supergravity theories 
has been constructed in four dimensions \cite{EMSS} 
and in five dimensions \cite{AFNS}. 
We need to consider topological defects such as 
junctions of walls, to consider a fundamental theory 
in space-time dimensions 
higher than five. 
The domain wall junctions 
have been studied \cite{AbrahamTownsend1}--\cite{NNS2} and 
can preserve a quarter of original SUSY. 
An exact analytic solution of the junction has been 
obtained in the ${\cal N}=1$ SUSY field 
theories in four dimensions \cite{OINS}. 
Possibility of junction solution has also been explored 
in supergravity \cite{CHT2}. 
The exact solution has been useful to unravel several unexpected 
 properties of domain wall junctions. 
The new Nambu-Goldstone fermion modes associated with the junction 
is found to be non-normalizable \cite{INOS}. 
The new central charge associated with the junction was once considered 
to be a mass of the junction. 
However the exact solution showed that the central charge 
contributes negatively to the energy of the junction 
\cite{OINS}, \cite{INOS}. 
Therefore it should more properly be interpreted as a 
binding energy of the walls which meet at the junction. 
As another toplogical defect with co-dimension two, 
an exact solution of vortices on $S^2$ has also been 
obtained before \cite{SakamotoTanimura}. 

The SUSY theories in dimensions higher than four are required 
to have at least eight supercharges. 
Theories with eight SUSY 
are often called ${\cal N}=2$ 
SUSY theories even in five or six dimensions, 
since they have twice as many SUSY charges 
compared to the simple SUSY theories in four dimensions. 
BPS wall solutions have been constructed in the 
${\cal N}=2$ SUSY nonlinear sigma models 
\cite{GPT}--\cite{Tong2}. 
Lump and Q-lump solutions preserving $1/8$ and $1/4$ SUSY, 
respectively, 
have also been considered 
\cite{AbrahamTownsend2}, \cite{NNS1}. 
On the other hand, the BPS wall junction has been 
constructed in linear \cite{OINS}, \cite{INOS} 
and nonlinear sigma models \cite{NNS2} only in 
${\cal N}=1$ SUSY models in four 
dimensions. 

The first analytic solution of the BPS junction has been 
obtained for an ${\cal N}=1$ $U(1)\times U(1)$ 
gauge theory with six charged and one neutral chiral scalar 
fields with minimal kinetic terms \cite{OINS}, 
which was constructed as a toy model for the 
${\cal N}=2$ $SU(2)$ gauge theory with one flavor 
\cite{SeibergWitten}. 
Subsequently it was realized 
that one can get rid of 
the vector multiplet by identifying six charged 
chiral scalar fields pair-wise into three 
chiral scalar fields. 
One still obtains the same junction solution as 
a BPS solution \cite{INOS} in this model with three 
``charged'' and one ``neutral'' 
chiral scalar fields with minimal kinetic terms 
(linear sigma model), without gauge field at all 
(Wess-Zumino model). 
It has also been shown that one can obtain the same solution 
in an ${\cal N}=1$ nonlinear sigma model 
with only single ``neutral'' chiral scalar field, 
by eliminating the other three ``charged'' 
chiral scalar fields 
appropriately \cite{NNS2}. 
In all these solutions, one finds that the ``neutral'' 
chiral scalar field plays a central role in constructing 
the junction solution. 
On the other hand, 
a neutral scalar field is contained in the ${\cal N}=2$ 
vector multiplet in the case of the ${\cal N}=2$ SUSY QED. 
Therefore it is tempting to embed the ${\cal N}=1$ 
gauge theory and its junction solution 
into the ${\cal N}=2$ SUSY QED.

The purpose of this paper is to give an exact analytic 
solution for 
the BPS domain wall junction in an ${\cal N}=2$ SUSY QED 
with three massive hypermultiplets. 
This is the first example of an exact junction solution 
in ${\cal N}=2$ SUSY theories. 
By explicitly working out eight SUSY transformations, we show 
that the junction solution 
 preserves two out of eight SUSY, namely it is a 
 ${1 \over 4}$ BPS state. 
Although the solution have many similarities with the 
previously obtained 
${1 \over 4}$ BPS junction solution in ${\cal N}=1$ SUSY 
theory, 
the resulting spectrum of the low-energy effective theory 
is richer. 
 For instance, we observe that there are zero modes 
 corresponding to 
spontaneously broken $U(1)$ global symmetries 
\cite{ShifmanYung}, \cite{Tong2}. 
Similarly to our previous solution in ${\cal N}=1$ 
theory\cite{OINS}, 
the Nambu-Goldstone modes on the junction background are 
not normalizable. 
As pointed out in Ref.\cite{INOS}, it may be possible 
to obtain 
a normalizable wave function when it is embedded into 
supergravity as explored in Ref.\cite{CHT2}. 
We also show that the same eight SUSY transformations 
can be derived from a nontrivial dimensional reduction of 
the ${\cal N}=2$ 
SUSY QED in five dimensions. 

The ${\cal N}=2$ SUSY theories with vector and 
hypermultiplets 
were introduced by Fayet using an automorphism of SUSY 
algebra \cite{Fayet}. 
He used both ${\cal N}=1$ superfield and component 
formalisms. 
The ${\cal N}=1$ superfield formalism makes only 
four SUSY manifest, 
but has been useful also to write down 
massless nonlinear sigma models 
\cite{LR}. 
Harmonic superspace formalism can make all eight SUSY 
manifest \cite{Ivanov}--\cite{ivanov-EH} 
and has been used to formulate ${1 \over 2}$ BPS 
equations to obtain BPS walls \cite{ANNS}, \cite{Zupnik2}. 
Even in the harmonic superspace formalism, however, 
it has been useful to use the Wess-Zumino gauge to 
clarify the physical field content of the theory 
 \cite{ANNS}. 
The Wess-Zumino gauge in the component formalism 
allows us to construct all the eight SUSY transformations 
explicitly. 
We also find that the action in terms of component fields 
can be assembled into ${\cal N}=1$ superfield formalisms 
making four out of eight SUSY manifest in two ways. 
Namely we can rewrite the same action in terms of two 
different superfields. 
One of them makes a set of four SUSY manifest, and the other 
makes the set of remaining four SUSY manifest. 
Of course we cannot make eight SUSY manifest in any 
one of the ${\cal N}=1$ superfield formalisms. 
We shall here employ the ${\cal N}=1$ superfield 
formalism \cite{MirabelliPeskin}, \cite{AGW}--\cite{LLP} 
as well as the component formalism both in the 
Wess-Zumino gauge. 

We find it essential to allow complex mass parameters in 
order to obtain a junction solution. 
The ${\cal N}=2$ SUSY theories are often derivable by 
means of a dimensional reduction from five and/or six 
dimensions \cite{SierraTownsend}. 
In this spirit, we also show that these ${\cal N}=2$ 
SUSY transformations can be understood 
in terms of a massive ${\cal N}=2$ theory in five dimensions. 
Since the massive theory in five dimensions can be obtained 
by a nontrivial dimensional reduction \`a la 
Scherk and Schwarz \cite{ScherkSchwarz} in one spacial 
direction, the mass parameter should be real. 
Therefore we find that it is difficult to extend 
our junction solution in the eight SUSY theory 
to a junction solution of ${\cal N}=2$ SUSY theory 
in five or six dimensions within the context of 
our multi-flavor QED. 
If we make a nontrivial dimensional reduction for 
 two spacial directions from six dimensions, 
we can obtain complex mass parameters. 
Therefore ${\cal N}=2$ SUSY theories in four dimensions 
can have complex mass parameters 
which allow the junction solutions.

In  sect.~\ref{sc:min_energyBPS}, our model of 
${\cal N}=2$ SUSY  massive multiflavor 
QED is introduced and 
BPS equations are derived as a minimum energy condition,  
and are shown to conserve one out of four SUSY 
in the ${\cal N}=1$ superfield formalism. 
In sect.~\ref{sc:junction}, 
an exact junction solution 
is obtained as a solution of ${1 \over 4}$ BPS 
equations of ${\cal N}=1$ superfield formalism. 
Zero modes are also briefly analyzed. 
In sect.~\ref{sc:8susy}, the remaining SUSY transformations 
are found by means of an automorphism of 
SUSY algebra. 
We also show that our model is invariant under the eight SUSY 
transformations and our BPS junction solution preserves two out of 
eight SUSY. 
Sect.~\ref{sc:DimRed} is devoted to relate 
the eight SUSY transformations in four-dimensions 
from ${\cal N}=2$ SUSY transformations in five dimensions. 

%%%%%%%%%%%%%%%%%%%%%%%%%%%%%%%%%%%%%%
\section{${\cal N}=2$ SUSY QED and BPS equations 
}
\label{sc:min_energyBPS}
%%%%%%%%%%%%%%%%%%%%%%%%%%%%%%%%%%%%%%

As one of the simplest models with eight 
SUSY, we consider ${\cal N}=2$ SUSY model with 
local $U(1)$ gauge symmetry in four dimensions with 
the gauge coupling constant $g$. 
If an ${\cal N}=1$ SUSY vector superfield $V_+$ is combined 
with an ${\cal N}=1$ SUSY 
chiral scalar superfield $\Phi_+$, an 
 ${\cal N}=2$ SUSY vector multiplet is obtained. 
In order to distinguish the four SUSY from the remaining 
four SUSY 
which will appear later, we denote the ${\cal N}=1$ 
superfield 
here by a suffix $+$. 
Combining ${\cal N}=1$ SUSY 
 chiral scalar superfields $Q_{+a}$ with $U(1)$ charge $+1$, 
and $\tilde Q_{+a}$ with $U(1)$ charge $-1$ gives 
an ${\cal N}=2$ hypermultiplet. 
The suffix $a=1, \dots, n$ denotes flavor. 
The ${\cal N}=2$ SUSY allows us to introduce the mass $m_a$ 
of the hypermultiplet for each flavor. 
Since our gauge symmetry is $U(1)$, the electric 
$c \in {\bf R}$ 
and the magnetic $b \in {\bf C}$ FI parameters 
 can also be introduced without violating 
the ${\cal N}=2$ SUSY \cite{Fayet}. %, \cite{APT}. 
Assuming a minimal kinetic term for the ${\cal N}=2$ 
vector and hypermultiplets, we thus obtain the ${\cal N}=2$ 
SUSY massive multiflavor QED. 
Using ${\cal N}=1$ superfield formalism, the Lagrangian is 
given by\footnote{
We use mostly the conventions of Wess and Bagger\cite{WB} 
for the ${\cal N}=1$ superfields, spinor and other 
notations.} 
\begin{eqnarray}
{\cal L}
&\!\!\!\!=&\!\!\!\!
\frac{1}{4g^2}\Bigl(W_+^\alpha W^+_\alpha 
\Big|_{\theta_+^2}+\bar{W}^+_{\dot \alpha}
\bar{W}_+^{\dot \alpha}
\Big|_{\bar{\theta}_+^2}\Bigr)
+\frac{1}{2g^2}\Phi_+^\dagger\Phi_+ 
\Big|_{\theta_+^2\bar{\theta}_+^2}
+\sum_{a=1}^n\Bigl(Q^\dagger_{+a}{\rm e}^{2V_+}Q_{+a}
+\tilde{Q}^\dagger_{+a}{\rm e}^{-2V_+}\tilde{Q}_{+a}\Bigr)
\Big|_{\theta_+^2\bar{\theta}_+^2} 
\nonumber \\
&\!\!\!\! &\!\!\!\!
-2c V_+ \Big|_{\theta_+^2\bar{\theta}_+^2}
+\left(\sum_{a=1}^n\Bigl(\Phi_+ -m_a\Bigr)Q_{+a}\tilde{Q}_{+a}
\Big|_{\theta_+^2}
-b\,\Phi_+ \Big|_{\theta_+^2}
+\textrm{h.c.}\right)
 \, , \label{lag1}
\end{eqnarray}
where the ${\cal N}=1$ vector multiplet $V_+$ and the 
chiral scalar 
multiplet $\Phi_+$ are multiplied by the gauge coupling $g$ 
to make the 
${\cal N}=2$ SUSY more easily visible. 
The coupling of $\Phi_+$ with the hypermultiplets 
$Q_{+a}, \tilde Q_{+a}$ 
in the last line of Eq.(\ref{lag1}) is dictated 
by the requirement of the ${\cal N}=2$ SUSY. 
If the mass parameters are absent $m_a=0$, the Lagrangian is 
invariant under the following global $U(n)$ transformations: 
\begin{equation}
{Q}_{+a} \rightarrow {Q}_{+a}'={Q}_{+b} g_{ba} , \qquad 
\tilde{Q}_{+a} \rightarrow 
\tilde{Q}_{+a}'=(g^\dagger)_{ab} \tilde{Q}_{+b}, 
\qquad \Phi_+ \rightarrow \Phi_+, \quad 
 V_+ \rightarrow V_+, \qquad g \in U(n). 
\end{equation}
The subgroup $U(1)$ of $U(n)=U(1)\times SU(n)$ is gauged. 
The mass parameters $m_a$ break the remaining global symmetry 
$SU(n)$ to $U(1)^{n-1}$. 
If $b=0$ in addition to $m_a=0$, 
the ${\cal N}=1$ superfield Lagrangian (\ref{lag1}) 
appears to have another global $U(1)$ symmetry : 
\begin{equation}
{Q}_{+a} \rightarrow {Q}_{+a}'={\rm e}^{i\alpha}{Q}_{+a}, 
\qquad 
\tilde{Q}_{+a} \rightarrow 
\tilde{Q}_{+a}'={\rm e}^{i\beta}\tilde{Q}_{+a}, 
\qquad \Phi_+ \rightarrow 
\Phi'_+= {\rm e}^{-i\beta-i\alpha}\Phi_+, 
%\quad  V \rightarrow V. 
\end{equation}
with $V_+$ invariant. 
This invariance respects ${\cal N}=1$, but 
is inconsistent with the ${\cal N}=2$ SUSY, since the chiral 
scalar field $\Phi_+$ should have the same transformation 
as the vector multiplet $V_+$ to form an ${\cal N}=2$ 
vector multiplet. 
Summarizing, our model with generic values of $m_a$ 
has the following $U(1)^n$ symmetries, which are 
consistent with the ${\cal N}=2$ SUSY : 
\begin{eqnarray}
Q_{+a} 
\rightarrow 
{\rm e}^{i\alpha_a} Q_{+a}, 
\quad 
\tilde{Q}_{+a}
\rightarrow 
{\rm e}^{-i\alpha_a}\tilde{Q}_{+a}, 
\quad 
\Phi_+ 
\rightarrow 
\Phi_+, 
\quad 
V_+ 
\rightarrow 
V_+. 
\label{eq:U1n-symmetry}
\end{eqnarray}
The diagonal $U(1)$ ($\alpha_1=\cdots=\alpha_n$) 
is a local gauged symmetry. 
Other $U(1)^{n-1}$ groups constrained by 
$\sum_{a=1}^n \alpha_a=0$ are global symmetries. 

To make the physical content of the theory more transparent, 
we shall use the Wess-Zumino gauge for the ${\cal N}=1$ 
vector 
superfield $V_+$. 
Then the ${\cal N}=1$ vector superfields can be expanded 
in terms of Grassmann number 
$\theta_+$ into component fields 
\begin{equation}
V_+(x,\theta_+, \bar \theta_+)
=
-\theta_+\sigma^m\bar{\theta}_+ v_m(x) 
+i\theta_+^2\bar{\theta}_+\bar{\lambda}(x)
-i\bar{\theta}_+^2\theta_+\lambda(x) 
+\frac{1}{2}\theta_+^2\bar{\theta}_+^2X_3(x) , 
\label{eq:vector1L}
\end{equation}
where $v_m$, $\lambda$, and $X_3$ are gauge field, gaugino, 
and auxiliary field, respectively. 
The ${\cal N}=1$ chiral scalar superfields can also be 
expanded into 
components using $y^m=x^m+i\theta_+\sigma^m\bar\theta_+$ 
as usual \cite{WB}
\begin{eqnarray}
\Phi_+(y, \theta_+)
&\!\!=&\!\!
\phi(y)+\sqrt{2}\theta_+(-i\sqrt{2}\psi(y) )+\theta_+^2(X_1(y)+iX_2(y)) 
\label{eq:vector2L}\\
Q_{+a}(y, \theta_+)
&\!\!=&\!\!
q_a(y)+\sqrt{2}\theta_+\psi_{q_a}(y)
+\theta_+^2F_a(y) 
\label{eq:hyper1L}\\
\tilde{Q}_{+a}(y, \theta_+)
&\!\!=&\!\!
\tilde{q}_a(y)+\sqrt{2}\theta_+\psi_{\tilde{q}_a}(y)
+\theta_+^2\tilde{F}_a(y) \, . 
\label{eq:hyper2L}
\end{eqnarray}
where the scalar fields are denoted by a small letter 
corresponding to 
the superfields, such as positively charged scalar 
$q_a$ as the first 
component of the superfield $Q_{+a}$. 
Let us note that the suffix $+$ is not carried by 
component fields, 
but is carried only by superfields, which are the 
functions of the 
associated Grassmann number $\theta_+$. 

In terms of component fields, the bosonic part of this 
Lagrangian becomes
\begin{eqnarray}
{\cal L}_{\rm boson}\!\!&=&\!\!
-\frac{1}{4g^2}v_{mn}v^{mn}
+\frac{1}{2g^2}(X_3)^2
-\frac{1}{2g^2}\left|\partial_m\phi\right| 
+\frac{1}{2g^2}|X_1+iX_2|^2
\nn \\
\!\!&&\!
+\sum_{a=1}^n\Bigl[|F_a|^2+|\tilde{F}_a|^2
-\left|{\cal D}_m q_a\right|^2
-\left|{\cal D}_m \tilde{q}_a\right|^2
+X_3(q^*_aq_a-\tilde{q}^*_a\tilde{q}_a)\Bigr]-c X_3 \nn \\
\!\!&&\!
+\sum_{a=1}^n\Bigl[(\phi-m_a)q_a\tilde{F}_a
+(\phi-m_a)F_a\tilde{q}_a
+(X_1+iX_2)q_a\tilde{q}_a\Bigr] -b\Bigl(X_1+iX_2\Bigr) \nn \\
\!\!&&\!
+\sum_{a=1}^n\Bigl[(\phi^* -m_a^* )q^*_a\tilde{F}_a^* 
+(\phi^* -m_a^* )F_a^*\tilde{q}_a^*
+(X_1-iX_2) q_a^*\tilde{q}_a^* \Bigr]-b^* 
\Bigl(X_1-iX_2\Bigr), 
\label{lag2}
\end{eqnarray}
where 
the field strength $v_{mn}$ 
and the covariant derivatives ${\cal D}_m$ are defined by 
\begin{equation}
v_{mn}=\partial_m v_n -\partial_n v_m, 
\qquad 
{\cal D}_m q_a \equiv (\partial_m  + i v_m)q_a,  \qquad 
{\cal D}_m \tilde q_a \equiv (\partial_m  - i v_m)
\tilde q_a , 
\end{equation}
respectively. 
The entire Lagrangian including the fermions will be given in 
sect.\ref{sc:8susy} where the full ${\cal N}=2$ SUSY will 
be clarified. 
We see that scalar field $q_a$ with the $U(1)$ charge $+1$ 
and $\tilde{q_a}$ with charge $-1$ have a complex mass $m_a$. 
Since a complex mass common to all the flavors can be 
absorbed 
by shifting the neutral complex scalar field $\phi$, 
these mass parameters 
can always be chosen to satisfy 
\begin{equation}
\Sigma_{a=1}^{n} m_a=0. 
\end{equation}
The real FI parameter $c$ of the $D$-term is usually 
called the electric FI parameter, and the complex parameter 
 $b$ appearing in the F-term is called the magnetic 
FI parameter \cite{Fayet}. %, \cite{APT}. 

The SUSY auxiliary fields $X_1, X_2, X_3, 
F_a,\tilde F_a$ can be eliminated by solving their algebraic 
equations of motion 
\begin{eqnarray}
X_3&=&-g^2\{\sum_{a=1}^n(|q_a|^2-|\tilde{q}_a|^2)-c\} 
\label{eom1}\\
X_1+iX_2&=&-2g^2(\sum_{a=1}^n q^*_a\tilde{q}_a^* -b^* ) 
\label{eom2} \\
F_a&=&-(\phi^* -m_a^* )\tilde{q}_a^* \label{eom3} \\
\tilde{F}_a&=&-(\phi^* -m_a^* )q^*_a \, . \label{eom4}
\end{eqnarray}
Then, the Lagrangian is given entirely in terms of 
physical fields 
\begin{eqnarray}
{\cal L}_{\rm boson}&=&-\frac{1}{4g^2}v_{mn}v^{mn}
-\frac{1}{2g^2}|\,
\partial\phi \,|^2-\sum_{a=1}^n 
\Bigl(|{\cal D}q_a|^2+|{\cal D}\tilde{q}_a|^2\Bigr)
-2g^2|\sum_{a=1}^n q_a\tilde{q}_a-b\,|^2 \nn \\
& &-\sum_{a=1}^n |\phi -m_a|^2
\Bigl(|q_a|^2+|\tilde{q}_a|^2\Bigr)-\frac{g^2}{2}
\Bigl\{\sum_{a=1}^n (|q_a|^2-|\tilde{q}_a|^2)-c\Bigr\}^2 
\, . \label{lag3}
\end{eqnarray}
A similar model has been considered 
previously in a different context 
\cite{Tong}--%, \cite{ShifmanYung}, 
\cite{Tong2}. 

SUSY vacua are given by vanishing auxiliary fields 
: $X_1=X_2=X_3=0$, 
and 
$F_a=\tilde F_a=0$. 
In the generic case of distinct complex mass parameters 
$m_a\neq m_b$ for $a \not=b$, 
we find precisely $n$ isolated SUSY vacua 
(and no other vacua). 
We denote the modulus and phase of the magnetic FI parameter 
$b$ by two real parameters $h>0$ and $\beta$ as 
\begin{equation}
b=h^2{\rm e}^{i\beta} . 
\label{eq:phase_b}
\end{equation}
The $i$-th vacuum is characterized by nonvanishing values of 
$q_a$, $\tilde q_a$ and $\phi$ : 
\begin{eqnarray}
\phi = m_a,
\qquad 
q_a=\sqrt{\sqrt{c^2 +4h^4} +c \over 2}{\rm e}^{i\alpha_a} 
\quad \tilde{q}_a=
\sqrt{\sqrt{c^2 +4h^4} -c \over 2}
{\rm e}^{i(\beta -\alpha_a )}, 
\label{vac1}
\end{eqnarray}
with vanishing values for the remaining hypermultiplets 
$q^*_b=\tilde{q}^*_b=0 \quad (a\neq b) $. 
At the $a$-th vacuum, the phase $\alpha_a$ is fixed 
breaking a $U(1)$ symmetry which is a linear combination 
of local gauged $U(1)$ and other global $U(1)^{n-1}$ 
generators. 
Because of Higgs mechanism, gauge boson should become 
massive in the vacuum. 
However, there still remains $U(1)^{n-1}$ global 
symmetries $\alpha_b, \; b\not=a$ unbroken as given in 
Eq.(\ref{eq:U1n-symmetry}). 

 The Hamiltonian corresponding to the 
Lagrangian (\ref{lag3}) is given by 
\begin{eqnarray}
{\cal H}
&=&
\frac{1}{2g^2}(v_{01}^2+v_{02}^2+v_{03}^2+v_{12}^2
+v_{13}^2+v_{23}^2) 
+\frac{1}{2g^2}\Bigl[|\partial_{0}\phi |^2
+|\partial_{1}\phi |^2+|\partial_{2}\phi |^2
+|\partial_{3}\phi |^2\Bigr] \nn \\
& &+\sum_{a=1}^n
\Bigl[|{\cal D}_{0}q_a|^2+|{\cal D}_{1}q_a |^2 
+|{\cal D}_{2}q_a|^2 +|{\cal D}_{3}q_a|^2 
+|{\cal D}_{0}\tilde{q}_a|^2+|{\cal D}_{1}\tilde{q}_a|^2 
+|{\cal D}_{2}\tilde{q}_a|^2 
+|{\cal D}_{3}\tilde{q}_a|^2 \Bigr] \nn \\
& &+\sum_{a=1}^n 
|\phi -m_a|^2\Bigl(|q_a|^2+|\tilde{q}_a|^2\Bigr) 
+\frac{g^2}{2}\{\sum_{a=1}^n 
(|q_a|^2-|\tilde{q}_a|^2)-c\}^2
+2g^2|\sum_{a=1}^n q_a\tilde{q}_a-b|^2 . 
\end{eqnarray}
Since the static domain wall junctions has nontrivial 
dependence only in 
two-dimensional spatial coordinates, 
we shall look for field configurations as a function of 
 $x^1$ and $x^2$ coordinates and introduce 
complex coordinates 
$z=x^1+ix^2$, $\bar{z}=x^1-ix^2$, 
$\partial_z=\frac{1}{2}(\partial_{1}-i\partial_{2})$, 
and $\partial_{\bar{z}}=\frac{1}{2}
(\partial_{1}+i\partial_{2})$. 
We also wish to maintain $1+1$ dimensional 
Lorentz invariance in the $x^0, x^3$ plane. 
Therefore we need to require 
\begin{equation}
v_0=v_3=0,  \qquad 
v_{01}=v_{02}=v_{03}=v_{13}=v_{23}=0 , \label{BPS5} 
\end{equation}
\begin{equation}
\partial_{0}\phi=0, \qquad {\cal D}_{0}q_a=0, 
\qquad {\cal D}_{0}\tilde{q}_a=0, \label{BPS6}
\end{equation}
\begin{equation}
\partial_{3}\phi =0, \qquad {\cal D}_{3}q_a =0, 
\qquad {\cal D}_{3}\tilde{q}_a =0 \, . \label{BPS7}
\end{equation}
In order to find a minimum energy configuration for a given 
boundary condition, we form complete squares \cite{CHT}, 
\cite{OINS} in the 
energy density functional 
${\cal E}$ by introducing an arbitrary phase 
$\Omega, \; |\Omega|=1$ 
\begin{eqnarray}
&
&
{\cal E}
=
\frac{1}{2g^2}\Bigl[v_{12}
+g^2\{\sum_{a=1}^n (|q_a|^2-|\tilde{q}_a|^2)-c\}\Bigr]^2 
+\frac{2}{g^2}\Big|\partial_{z}\phi 
-g^2\Omega(\sum_{a=1}^n q_a^*\tilde{q}_a^* -b^* )\Big|^2 
\nn \\
&&+4\sum_{a=1}^n \Big|{\cal D}_{z}q_a
-\frac{1}{2}\Omega(\phi^* -m_a^* )\tilde{q}_a^*\Big|^2 
+4\sum_{a=1}^n \Big|{\cal D}_{z}\tilde{q}_a
-\frac{1}{2}\Omega (\phi^* -m_a^* )q_a^* \Big|^2 \nn \\
&&
+cv_{12} 
+\partial_{z}(2\Omega^*(\sum_{a=1}^n 
(\phi -m_a)q_a\tilde{q}_a -b\phi))
+\partial_{\bar{z}}(2\Omega(\sum_{a=1}^n (\phi^* -m_a^* )
q_a^*\tilde{q}_a^* -b^*\phi^* ))
\nn \\
&&
+\sum_{a=1}^n \Bigl[\partial_{z}(q_a^*{\cal D}_{\bar{z}}q_a 
-q_a({\cal D}_{z}q_a)^* )
+\partial_{z}(\tilde{q}_a^*{\cal D}_{\bar{z}}\tilde{q}_a 
-\tilde{q}_a({\cal D}_{z}\tilde{q}_a)^* )\Bigr] \nn \\
&&
+\sum_{a=1}^n \Bigl[\partial_{\bar{z}}
(q_a({\cal D}_{\bar{z}}q_a)^* 
-q_a^*{\cal D}_{z}q_a ) 
+\partial_{\bar{z}}(\tilde{q}_a
(\tilde{{\cal D}}_{\bar{z}}\tilde{q}_a)^* 
-\tilde{q}_a^*{\cal D}_{z}\tilde{q}_a )\Bigr] 
\nn \\
&&
+\frac{1}{2g^2}\partial_{z}(\phi^*\partial_{\bar{z}}\phi 
-\phi\partial_{\bar{z}}\phi^* ) 
+\frac{1}{2g^2}\partial_{\bar{z}}(\phi\partial_{z}\phi^* 
-\phi^*\partial_{z}\phi ) 
 . \label{energy}
\end{eqnarray}

The last four lines are total derivatives 
that give surface terms when integrated over 
 the entire $x^1, x^2$ plane. 
Since all 
the remaining terms are the complete squares, 
we find that the integrated energy over $x^1, x^2$ plane
of the field configuration is always larger than the surface 
terms, which 
are completely determined by the 
boundary condition at spatial infinity. 
This bound is called the BPS bound and is saturated by 
requiring the 
complete squares to vanish 
\begin{eqnarray}
&&v_{12}=
-g^2\left\{\sum_{a=1}^n (|q_a|^2-|\tilde{q}_a|^2)-c\right\} 
\label{BPS4}
\\
&&{1 \over g^2}
{\partial \phi \over \partial z} 
=\Omega\left(\sum_{a=1}^n q_a^*\tilde{q}_a^* -b^* \right) 
\label{BPS1}\\
&&2{\cal D}_{z}q_a=
%\frac{1}{2}
\Omega(\phi^* -m_a^* )\tilde{q}_a^*  
\label{BPS2}\\
&&2{\cal D}_{z}\tilde{q}_a=
%\frac{1}{2}
\Omega (\phi^* -m_a^* )q_a^* .
\label{BPS3}
\end{eqnarray}
 These first order differential equations are called 
 the BPS equations. 
Since the surface terms depend on $\Omega$, 
 the phase factor $\Omega$ 
can be chosen to obtain the best bound. 
Let us also note that the minimum energy configurations 
automatically 
satisfy the equations of motion \cite{CHT}, \cite{OINS}.

Since the Lagrangian (\ref{lag1}) with ${\cal N}=1$ 
 superfield exhibits the ${\cal N}=1$ SUSY manifestly, 
we can formulate the condition of partial conservation 
of SUSY. 
We will see that the above minimum energy conditions 
(\ref{BPS4})--(\ref{BPS3}) 
are precisely the conditions to conserve 
one out of four SUSY. 
We need to consider only the SUSY transformations of 
fermions, 
since only bosonic fields can have nonvanishing values. 
The ${\cal N}=1$ SUSY transformations of gaugino is given 
by \cite{WB} 
\begin{eqnarray}
\delta_{\xi_+} \lambda 
&=& 
\sigma^{mn}v_{mn}\xi_++iX_3\xi_+ \nonumber \\
&=& 
\left[
\begin{array}{cc}
v_{03}-iv_{12}+iX_3 & v_{01}+v_{13}-iv_{23}-iv_{02} \\
v_{01}-v_{13}-iv_{23}+iv_{02} & -v_{03}+iv_{12}+iX_3 
\end{array}
\right] 
\left[
\begin{array}{c}
\xi_{+1} \\
\xi_{+2} 
\end{array}
\right] 
\, . 
\end{eqnarray}
If we require that a part of SUSY corresponding to the 
upper component $\xi_{+1}$ is conserved ($\xi_{+2}=0$), 
we find \cite{OINS}  
\begin{eqnarray}
v_{12}
=
X_3, \quad  v_{03}=0 , 
\quad 
v_{01}
=
v_{13},\quad v_{23}=v_{02} \, . 
\label{PSC5}
\end{eqnarray}
Using the algebraic equation of 
motion for the auxiliary field (\ref{eom1}), 
the minimum energy condition (\ref{BPS4}) for the 
vector multiplet 
is precisely the same as the condition of the 
partial SUSY conservation condition 
(\ref{PSC5}). 
Similarly the ${\cal N}=1$ SUSY transformations 
of fermions in chiral scalar multiplets 
are given by \cite{WB} 
\begin{eqnarray}
\delta_{\xi_+}(-i\sqrt{2}\psi )&=&
i\sqrt{2}\sigma^m\partial_m \left(\phi-m_a\right) \bar{\xi}_+
+\sqrt{2}(X_1+iX_2)\xi_+ , \\
\delta_{\xi_+}\psi_{q_a}&=&
i\sqrt{2}\sigma^m{\cal D}_m q_a\bar{\xi}_+
+\sqrt{2}F_a\xi_+ , 
\\
\delta_{\xi_+}\bar{\psi}_{\tilde{q}_a}&=&
i\sqrt{2}\bar{\sigma}^m{\cal D}_m\tilde{q}^*_a\xi_+
+\sqrt{2}\tilde{F}_a\bar{\xi}_+ . 
\end{eqnarray}
 For these transformations, 
we express derivatives in terms of complex 
coordinates, assuming $x_1$, $x_2$ dependence only 
\begin{eqnarray}
\sigma^m\partial_m&=&
(\sigma^1+i\sigma^2)\frac{1}{2}(\partial_1-i\partial_2)
+(\sigma^1-i\sigma^2)\frac{1}{2}(\partial_1+i\partial_2) , 
\nn \\
&=&2\sigma^{+}\partial_{z}+2\sigma^{-}\partial_{\bar{z}} \, ,
\end{eqnarray}
where
$\sigma^{+}\equiv (\sigma^1+i\sigma^2)/2$,
$\sigma^{-}\equiv (\sigma^1+i\sigma^2)/2$. 
If we require conservation of only one out of four SUSY 
specified by 
\begin{eqnarray}
&& -\Omega \sigma^{+}\bar{\xi}_+ = 
i\xi_+ \quad \textrm{and}\quad \sigma^{-}\bar{\xi}_+=0 , 
\label{eq:quarter1}
\end{eqnarray}
we obtain \cite{OINS}  
\begin{eqnarray}
\delta_{\xi_+}(-i\sqrt{2}\psi) &=&
i\sqrt{2}\sigma^m\partial_m\left(\phi-m_a\right)
\bar{\xi}_++\sqrt{2}(X_1+iX_2)\xi_+ , 
\nn \\
&=&
i2\sqrt{2}(\sigma^{+}\partial_z
+\sigma^{-}\partial_{\bar{z}})\left(\phi-m_a\right)\bar{\xi}_+
+\sqrt{2}(X_1+iX_2)\xi_+ , 
\nn \\
&=&
\sqrt{2}\xi_+\Omega^{-1}\left(2\partial_z
\left(\phi-m_a\right) 
+\Omega(X_1+iX_2)\right) .
\end{eqnarray}
\begin{eqnarray}
\delta_{\xi_+}\psi_{q_a}&=&
i\sqrt{2}\sigma^m{\cal D}_m q_a\bar{\xi}_+
+\sqrt{2}{F}_a\xi_+ , 
\nn \\
&=&
\sqrt{2}{\xi}_+\Omega^{-1}\left(2{\cal D}_z q_a 
+\Omega F_a \right) . 
\end{eqnarray}
\begin{eqnarray}
\delta_{\xi_+}\bar{\psi}_{\tilde{q}_a}&=&
i\sqrt{2}\bar{\sigma}^m{\cal D}_m\tilde{q}^*_a\xi_+
+\sqrt{2}\tilde{F}_a^\ast\bar{\xi}_+ , 
\nn \\
&=&
\sqrt{2}{\xi}_+\Omega^{-1}\left(2{\cal D}_z \tilde{q}_a 
+\Omega \tilde{F}_a \right) . 
\end{eqnarray}
Therefore we find that the condition of conservation of 
one out of four SUSY is given by 
\begin{equation}
{1 \over g^2}{\partial \phi \over \partial z} 
= 
-\Omega \left(X_1+iX_2\right), 
\quad 
2{\cal D}_{z} q_a 
=
-\Omega F_a, 
\quad 
2{\cal D}_{z} \tilde{q}_a 
=
-\Omega \tilde{F}_a
\label{eq:PSC_N1}
\end{equation}
The algebraic equations of motion for auxiliary fields are given 
in terms of the superpotential $P$ 
\begin{equation}
X_1+iX_2
= 
-\left({\partial P \over \partial \phi}\right)^*, 
\quad 
 F_a 
=
-\left({\partial P \over \partial q_a}\right)^*, 
\quad 
 \tilde{F}_a 
=
-\left({\partial P \over \partial \tilde{q}_a}\right)^*
\label{eq:aux_N1}
\end{equation}
This superpotential $P$ as a function of scalar fields 
is given in our case by
\begin{eqnarray}
P&\!\!=&\!\!\sum_{a=1}^n\Bigl(\phi -m_a\Bigr)q_a\tilde{q}_a-b\phi  \, . 
\label{potential}
\end{eqnarray}
Using the superpotential (\ref{potential}) and the algebraic 
equations of motion for auxiliary fields (\ref{eq:aux_N1}), 
we see that the minimum energy conditions 
(\ref{BPS1})--(\ref{BPS3}) are precisely the same 
as the conditions for the conservation of one out of four 
SUSY (\ref{eq:PSC_N1}).

%%%%%%%%%%%%%%%%%%%%%%%%%%%%%%%%%%%%%%
\section{Domain wall junction
}
\label{sc:junction}
%%%%%%%%%%%%%%%%%%%%%%%%%%%%%%%%%%%%%%

\begin{figure}[hbtp]
\begin{minipage}{.45\linewidth}\begin{center}
\scalebox{1.0}{\includegraphics{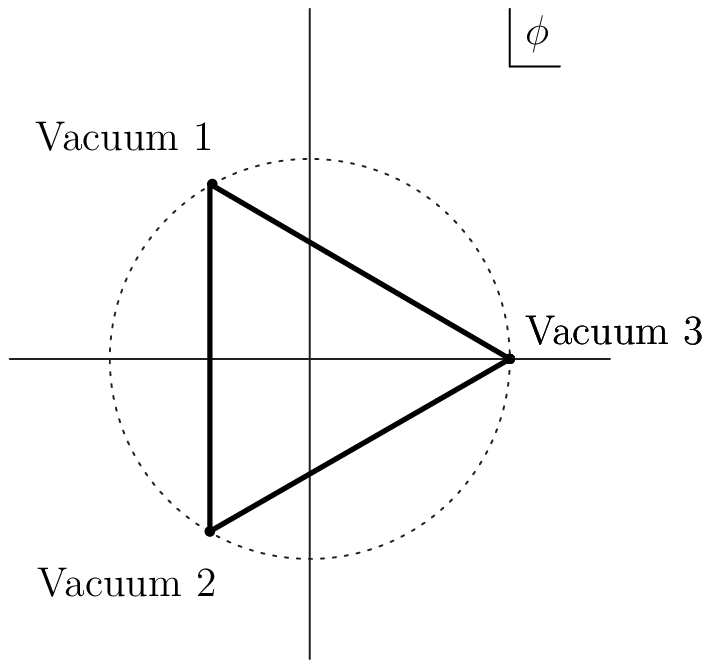}}
\caption{Vacua in complex $\phi$ plane. 
}
\label{vaca}
\end{center}
\end{minipage}
\begin{minipage}{.45\linewidth}
\begin{center}
	\scalebox{1.0}{\includegraphics{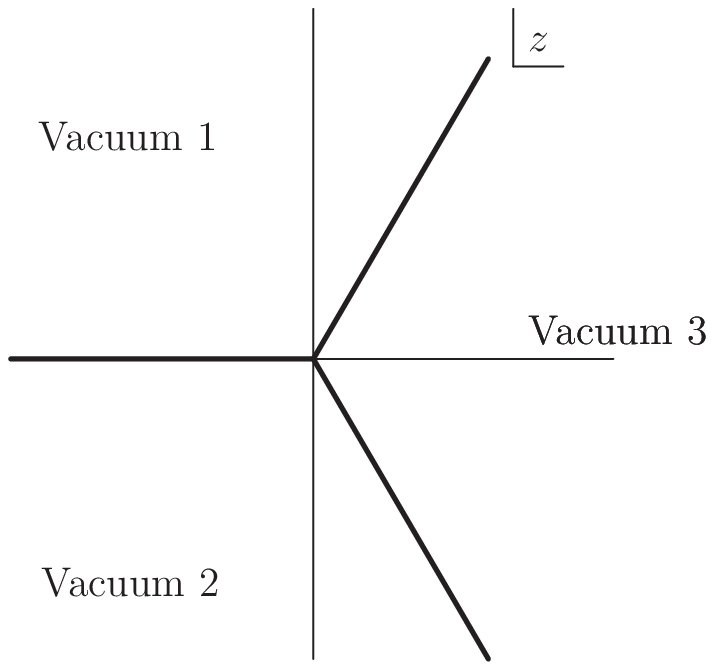}}
	\caption{The $Z_3$ junction in real space $z=x^1+ix^2$. 
}
	\label{vacb}
\end{center}
\end{minipage}
\end{figure}

In order to obtain an exact solution of junctions, we shall 
embed the known solution to a solution of BPS equations 
 (\ref{BPS4})--(\ref{BPS3}) for  our 
${\cal N}=2$ SUSY massive multiflavor QED. 
We are making one complex structure out of three manifest 
by using the ${\cal N}=1$ superfield formalism. 
Although three FI parameters 
$c$, $b_1={\rm Re} \, b$, and $b_2={\rm Im} \, b$ 
form an $SU(2)_R$ triplet, the choice of particular complex structure 
made the $SU(2)_R$ symmetry not visible. 
In this circumstance, we find it convenient to choose 
the FI parameters $c$ and $b$ in Eq.(\ref{eq:phase_b}) as 
\begin{equation}
c=0, 
\qquad 
b = h^2 \in {\bf R} \quad (\beta=0)
. 
\label{eq:bChoice}
\end{equation}
Then the $a$-th vacuum values of fields in Eq.(\ref{vac1}) become 
\begin{eqnarray}
\phi = m_a,
\qquad 
q_b= h {\rm e}^{i\alpha_a}\delta_{ba}, 
\quad \tilde{q}_b=h {\rm e}^{-i\alpha_a}\delta_{ba}. 
\label{eq:vac_b}
\end{eqnarray}
Since the known junction solution was obtained \cite{OINS} 
with the $Z_3$ symmetry for SUSY vacua 
and with a relation between 
the vacuum values of charged chiral scalar fields $q_a, \tilde q_a$ 
and the neutral scalar fields $\phi$, 
we should require  $n=3$ flavors with $Z_3$ symmetry, 
and a relation between the mass scales of 
$m_b$ and the Fayet-Iliopoulos term $h$. 
Altogether we assume the following 
particular values for the parameters of our model : 
\begin{equation}
m_b=\frac{2gh}{\sqrt{3}}{\rm e}^{i\frac{2\pi}{3}b},\qquad b=1,2,3  \, . 
\label{vac3}
\end{equation}
The resulting vacua are 
illustrated in the complex $\phi$ plane in Fig.\ref{vaca}. 

Combined with the algebraic 
equation of motion for auxiliary field $X_3$ (\ref{eom1}) 
for vector superfield, 
we can satisfy the BPS equations (\ref{PSC5}) 
trivially by choosing 
\begin{equation}
v_0(x^1, x^2)=v_3(x^1, x^2) =0
, 
\qquad 
|q_b(x^1, x^2)|=|\tilde q_b(x^1, x^2)| .
\label{eq:vector_D_cond1}
\end{equation}
Suggested by this condition, we assume the following relation 
between values of hypermultiplets \cite{OINS} 
\begin{equation}
q_b(x^1, x^2){\rm e}^{-i\alpha_b}=\tilde q_b(x^1, x^2){\rm e}^{i\alpha_b} 
\in {\bf R} ,
\label{eq:vector_D_cond2}
\end{equation}
in accordance with the vacuum values (\ref{eq:vac_b}) 
which should be reached at infinity. 
This assumption will be justified 
a posteriori after finding out solutions. 
We are interested in a BPS junction configuration separating 
three vacuum domains $a=1,2,3$ with $Z_3$-symmetry, 
where the third vacuum is placed 
at infinity along the positive real axis as illustrated 
in Fig.\ref{vacb}. 
This configuration corresponds to the choice of the phase factor 
$\Omega=-1$ \cite{OINS}, \cite{INOS}. 
Now the remaining BPS equations for chiral scalar multiplets 
(hypermultiplets and the chiral scalar $\phi$ in the ${\cal N}=2$ 
vector multiplet) read 
\begin{eqnarray}
&&2\frac{\partial q_b}{\partial z}
=
\left(\frac{2gh}{\sqrt{3}}{\rm e}^{i\frac{2\pi}{3}b}-\phi \right)^*
q_b , \label{sol1}\\
&&{1 \over g^2}\frac{\partial\phi}{\partial z}
=
h^2-
\sum_{b=1}^3|q_b|^2 \, . \label{sol2}
\end{eqnarray}

We define a dimensionless 
complex coordinate $\hat z$ and real dimensionless fields 
$\hat q_b$ and $\hat \phi$ by rescaling with the 
normalization factor associated to the vacuum values 
%$h{\rm e}^{i\alpha_b}$, and rescale the field $\phi$ 
as 
\begin{eqnarray}
z={\sqrt{3}\over 2}{1 \over gh }\hat{z}, 
\qquad \quad 
q_b =h{\rm e}^{i\alpha_b}\hat{q}_b ,
\qquad \quad \phi ={2 \over \sqrt{3}} g h \hat{\phi} 
\, . \label{sol3}
\end{eqnarray}
Then the BPS equations become 
\begin{eqnarray}
2\frac{\partial \hat{q}_b}{\partial \hat{z}}
&=&
\left(
{\rm e}^{i\frac{2\pi}{3}b}-\hat{\phi}
\right)^*
\hat{q}_b , 
\qquad \quad \hat{q_b} \in {\bf R} , 
\label{sol5} \\
2\frac{\partial \hat{\phi}}{\partial \hat{z}}
&=& 
{3 \over 2}
\Bigl(1-\sum_b \hat{q}_b^2\Bigr) \, . \label{sol4} 
\end{eqnarray}
We can now recognize the familiar form of the BPS equation allowing 
the junction as an exact solution \cite{OINS}, \cite{NNS2}. 
Let us define the following auxiliary quantities ${f}_b$ 
\begin{eqnarray}
f_b=\exp \left(\frac{1}{2}\left(
{\rm e}^{-i\frac{2\pi}{3}b}\hat{z}+
{\rm e}^{i\frac{2\pi}{3}b}\hat{z}^* \right)\right) ,
\qquad 
b=1,2,3
\, .  \label{sol6}
\end{eqnarray}
 Following identities can be derived for these auxiliary quantities 
 \cite{OINS}, \cite{NNS2}
\begin{eqnarray}
2 \frac{\partial}{\partial \hat{z}}\left(\frac{f_b}{f_1+f_2+f_3}\right)
=
\left({\rm e}^{-i\frac{2\pi}{3}b}
-\frac{\sum_c {\rm e}^{-i\frac{2\pi}{3}c}f_c}{f_1+f_2+f_3}\right)
\frac{f_b}{f_1+f_2+f_3} \, ,  \label{sol7}
\end{eqnarray}
\begin{eqnarray}
2\frac{\partial}{\partial \hat{z}}\left(\frac{\sum_b 
{\rm e}^{i\frac{2\pi}{3}b}f_b}{f_1+f_2+f_3} \right)
=
{3 \over 2}
\left(1-\frac{\sum_b f_b^2}{(f_1+f_2+f_3)^2} \right) \, .  \label{sol8}
\end{eqnarray}
Therefore we find 
the solutions for the BPS equations 
\begin{eqnarray}
\hat{q}_b
=
\frac{f_b}{f_1+f_2+f_3}, 
\qquad 
\hat{\phi} 
=
\frac{\sum_b 
{\rm e}^{i\frac{2\pi}{3}b}f_b}{f_1+f_2+f_3} \, .\label{sol10}
\end{eqnarray}
These solutions can be rewritten in terms of our original variables 
$q_b,\,\phi ,$ and $z$,
\begin{eqnarray}
q_b
&\!\!=&\!\!
h{\rm e}^{i\alpha_b}\frac{f_b}{f_1+f_2+f_3} 
 , \label{sol11}
\\
\tilde{q}_b&\!\!=&\!\!h{\rm e}^{-i\alpha_b}\frac{f_b}{f_1+f_2+f_3} , 
\\
\phi &\!\!=&\!\! {2 g h \over \sqrt{3}} 
\frac{\sum_b {\rm e}^{i\frac{2\pi}{3}b}f_b}{f_1+f_2+f_3} \, , \label{sol12}
\end{eqnarray}
where
\begin{eqnarray}
f_b=\exp \left(\frac{2gh}{\sqrt{3}}\frac{1}{2}
\left({\rm e}^{-i\frac{2\pi}{3}b}z
+{\rm e}^{i\frac{2\pi}{3}b}z^* \right)
\right) , \qquad
b=1,2,3
\, .  \label{sol13}
\end{eqnarray}

\begin{figure}[htbp]
\begin{center}
	\scalebox{.7}{\includegraphics{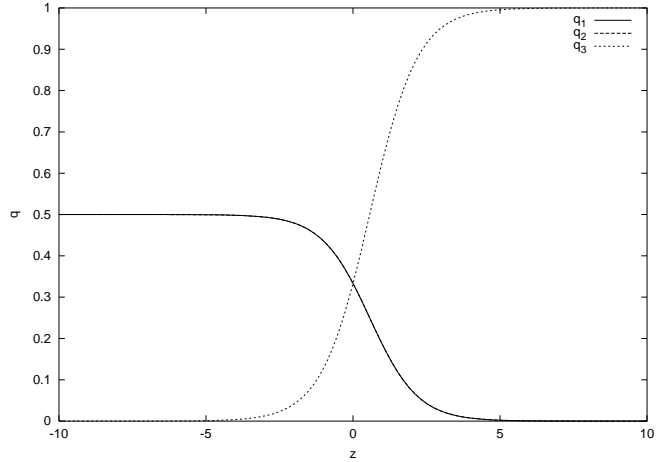}}
	\caption{
        Along $z\to +\infty$ one obtains the vacuum $3$, 
        where only $q_3$ takes non-zero values. 
        Along $z\to -\infty$ one obtains the middle point of wall between 
        first and second vacua, where $|q_1|=|q_2|$. }
	\label{q3}
\end{center}
\end{figure}

\begin{figure}[htbp]
\begin{center}
	\scalebox{0.9}{\includegraphics{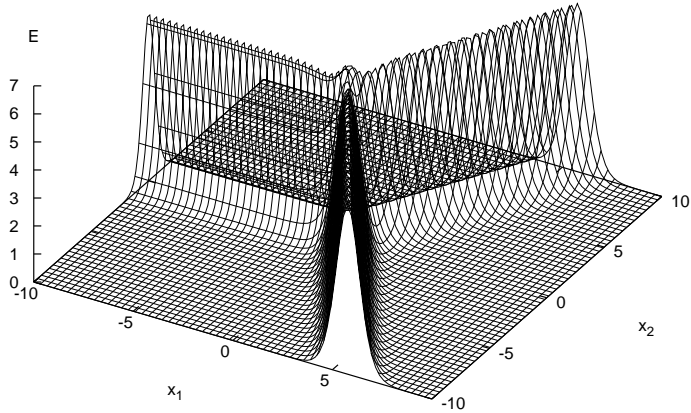}}
	\caption{Energy density of the $Z_3$ junction configuration 
        of ${\cal N}=2$ SUSY QED. }
	\label{8sj-energy}
\end{center}
\end{figure}

By rotating the field configuration by ${2\pi \over 3}$, 
we find that the solution (\ref{sol13}) 
is precisely the same field configuration 
of the previous junction solution in the ${\cal N}=1$ SUSY theory 
 \cite{OINS} 
provided the dimensionful parameter $h$ is related to the parameter 
$\Lambda$ through $h=\sqrt2 \Lambda$. 
The field $\phi$ of the junction configuration takes values 
inside the triangle connecting the three vacua as illustrated in 
Fig.\ref{vaca}. 
The straight line segments between vacua on the $\phi$ plane 
correspond to the spatial infinity in real space $z\rightarrow \infty$. 
As an illustration of the asymptotic behavior of the junction 
configuration, 
values of fields $q_a, \; a=1,2,3$ choosing 
$\alpha_a=0$ are shown 
along the real axis in Fig.\ref{q3}. 
The energy density computed analytically in our previous 
solution of ${\cal N}=1$ SUSY model \cite{OINS} can easily 
be converted into our case of ${\cal N}=2$ SUSY QED. 
The energy density of the junction solution in our 
${\cal N}=2$ SUSY model is shown in Fig.\ref{8sj-energy}.

Let us discuss zero modes on this junction solution. 
First we notice that we have two massless boson 
corresponding to spontaneously broken translation 
in two directions $x^1, x^2$. 
As for the two global $U(1)$ symmetries, 
both of them are broken on the junction solution, 
and there are two corresponding Nambu-Goldstone bosons. 
For each domain wall, only two hypermultiplets 
have nontrivial field configurations. 
Therefore only one of the two $U(1)$ global symmetries 
is spontaneously broken and only one Nambu-Goldstone 
boson associated with the $U(1)$ phase rotation appears 
\cite{ShifmanYung}, \cite{Tong2}. 
Since the broken $U(1)$ on each wall is a different 
combination of the two $U(1)$ global symmetries, 
the associated Nambu-Goldstone boson on each wall 
is also a different linear combination of these two. 
This situation is very similar to the property of the 
Nambu-Goldstone fermions on the junction 
that was observed previously 
\cite{OINS}, \cite{INOS}. 
As for fermions, six out of eight SUSY are 
spontaneously broken. 
Therefore we have six Nambu-Goldstone 
fermions. 
On each wall, only four SUSY are broken. 
Therefore there are four Nambu-Goldstone 
fermions on each wall. 
These four  Nambu-Goldstone 
fermions on each wall are different linear 
combinations of six Nambu-Goldstone fermions 
on the junction as a whole, since 
 different linear combinations of SUSY charges 
are broken on each wall. 
This situation is analogous to 
the previously obtained 
junction solution in ${\cal N}=1$ 
theory 
\cite{OINS}, \cite{INOS}. 
It has been pointed out that the wall provides a 
model to ``localize'' 
gauge bosons \cite{ShifmanYung}. 
In this context, one should note that 
one out of two Nambu-Goldstone bosons 
associated with the two $U(1)$'s mixes with 
the $U(1)$ gauge boson. 
We postpone a more detailed analysis of walls 
and junctions involving the 
gauge field for subsequent publications. 

Let us discuss possible junction configurations in 
other models. 
We have obtained an analytic solution of junction provided 
the mass parameters $m_i$ are tuned to the gauge coupling 
$g$ and the Fayet-Iliopoulos parameter $b=h^2$ as in 
Eq.(\ref{vac3}). 
However, it is almost clear from continuity that 
 junction configurations should 
exist even if the mass parameters 
are perturbed infinitesimally away form the tuned values. 
On the other hand, junction configuration is 
not allowed if masses are on a straight line in complex 
mass plane, for instance if all masses are aligned on the 
real axis. 
Since the tension is given by the absolute value of the 
difference of superpotential, the possible junction 
configuration cannot satisfy the condition of 
mechanical stability \cite{CHT} 
if masses are on a straight line. 
We conjecture that there should exist a ``line of marginal 
stability'' somewhere between the $Z_3$ symmetric 
mass parameters and mass parameters on a straight line. 
The junction should become unstable across the 
line of marginal stability. 
An interesting example of the line of 
marginal stability for domain walls 
has been explicitly demonstrated for 
a similar Wess-Zumino model \cite{PortuguesTownsend}. 
Lines of marginal stability for monopoles and dyons 
are well-studied in the ${\cal N}=2$ gauge theroies 
\cite{RSVV}. 

It is also likely that there exist junctions of $n$-walls, 
although it is difficult to work out explicit solutions 
except in particular nonlinear sigma models \cite{NNS2}. 

Similarly to our previous solution \cite{INOS}, 
the Nambu-Goldstone bosons and 
fermions on our junction solution are not normalizable. 
Therefore usual wisdom of a low-energy effective Lagrangian 
approach cannot be applied easily to our junction solution. 
However, it has been observed that a graviton zero mode 
is localized at the junction, if the bulk spacetime is warped 
\cite{ADDK}. 
This is due to a suppression factor produced by the 
bulk AdS(-like) spacetime. 
It may also be possible to exploit this mechanism to 
obtain normalizable zero modes localized at the junction. 
The general properties of possible 
junction solutions has been studied 
in the presence of gravity \cite{CHT2}, 
although no explicit solution has been obtained. 
This is an interesting subject in future.

%%%%%%%%%%%%%%%%%%%%%%%%%%%%%%%%%%%%%%%%
\section{8 SUSY transformations}
\label{sc:8susy}
%%%%%%%%%%%%%%%%%%%%%%%%%%%%%%%%%%%%%%%%

In this and the following sections, we return to the general 
case of  $n$-flavors. 
There have been a number of studies to formulate the 
${\cal N}=2$ SUSY field theories in five dimensions 
in terms of ${\cal N}=1$ 
superfield formalism \cite{MirabelliPeskin}, 
\cite{AGW}--\cite{LLP}. 
Inspired by these studies \cite{MirabelliPeskin}, \cite{AGW}, 
\cite{Hebecker}, 
 we will 
redefine auxiliary fields of chiral scalar fields for 
 a hypermultiplet (\ref{eq:hyper1L}) and (\ref{eq:hyper2L})
in order to identify all the eight supersymmetry transformations 
\begin{equation}
F_a=F^\prime_a-\left(\phi^*-m_a^*\right)\tilde{q}_a^*,
\qquad 
\tilde{F}_a=
-\tilde{F}^\prime_a-q_a^* \left(\phi^*-m_a^*\right)  
\end{equation}
Instead of Eqs.(\ref{eq:vector1L})-(\ref{eq:hyper2L}), 
the component expansions of superfields in powers of 
Grassmann number $\theta_+$ now read 
\begin{eqnarray}
V_+&=&
-\theta_+\sigma^m\bar{\theta}_+v_m 
+i\theta_+^2\bar{\theta}_+\bar{\lambda}
-i\bar{\theta}_+^2\theta_+\lambda 
+\frac{1}{2}\theta_+^2\bar{\theta}_+^2X_3 \label{vector1}\\
\Phi_+&=&
\phi+\sqrt{2}\theta_+(-i\sqrt{2}\psi )
+\theta_+^2(X_1+iX_2) \label{vector2}\\
Q_{+a}&=&
q_a+\sqrt{2}\theta_+\psi_{q_a}
+\theta_+^2(F^\prime_a-
\left(\phi^*-m_a^*\right)\tilde{q}_a^* ) 
\label{hyper1}\\
\tilde{Q}_{+a}&=&
\tilde{q}_a+\sqrt{2}\theta_+\psi_{\tilde{q}_a}
+\theta_+^2(-\tilde{F}^\prime_a
-q_a^* \left(\phi^*-m_a^*\right) )\, . 
\label{hyper2}
\end{eqnarray}

In terms of these component fields, the full Lagrangian 
(\ref{lag1}) 
is given by 
\begin{eqnarray}
{\cal L}&=&
-\frac{1}{4g^2}v_{mn}v^{mn}
-\frac{1}{g^2}i\lambda\sigma^m\partial_m\bar{\lambda}
+\frac{1}{2g^2}(X_3)^2 +\frac{1}{2g^2}|X_1+iX_2|^2 
\nn\\
&&
-\frac{1}{2g^2}|\partial_m \phi |^2
-\frac{1}{g^2}i\bar{\psi}\bar{\sigma}^m\partial_m\psi 
-cX_3-b(X_1+iX_2)-b^* (X_1-iX_2) 
\nn \\
&&
+\sum_{a=1}^n \Bigl[(F_a^{\prime*}-\tilde{q}_a \left(\phi-m_a\right))
(F^\prime_a-\left(\phi^*-m_a^*\right)\tilde{q}_a^*)-|{\cal D}_m q_a|^2 \nn \\
&&\qquad\quad
-i\bar{\psi}_{q_a}\bar{\sigma}^m{\cal D}_m\psi_{q_a} 
-i\sqrt{2}(\bar{\psi}_{q_a}\bar{\lambda}q_a
-\psi_{q_a}\lambda q_a^* )+X_3|q_a|^2 \Bigr] \nn \\
&&
+\sum_{a=1}^n \Bigl[(-\tilde{F}^\prime_a -q_a^* \left(\phi^*-m_a^*\right) )
(-\tilde{F}_a^{\prime*} -\left(\phi-m_a\right) q_a)
-|{\cal D}_m \tilde{q}_a|^2 \nn \\
&&\qquad\quad
-i\psi_{\tilde{q}_a}\sigma^m{\cal D}_m
\bar{\psi}_{\tilde{q}_a}
+i\sqrt{2}(\bar{\psi}_{\tilde{q}_a}
\bar{\lambda}\tilde{q}_a 
-\psi_{\tilde{q}_a}\lambda\tilde{q}_a^*)
-X_3|\tilde{q}_a|^2\Bigr] \nn \\
&&
+\sum_{a=1}^n\Bigl[(\phi -m_a)\{\tilde{q}_a 
(F^\prime_a-\left(\phi^*-m_a^*\right) \tilde{q}_a^* )
+ (-\tilde{F}^\prime_a -q_a^* \left(\phi^*-m_a^*\right) )q_a\} \nn \\
&&\qquad\quad
+(X_1+iX_2)\tilde{q}_aq_a -\psi_{\tilde{q}_a}
(\phi -m_a)\psi_{q_a} 
+i\sqrt{2}\psi_{\tilde{q}_a}\psi q_a 
+i\sqrt{2}\psi_{q_a}\psi \tilde{q}_a\Bigr] \nn \\
&&
+\sum_{a=1}^n\Bigl[(\phi^* -m_a^* ) 
\{ q_a^* (-\tilde{F}^{\prime*}_a-\left(\phi-m_a\right) q_a)
+(F_a^{\prime*} -\tilde{q}_a \left(\phi-m_a\right) )\tilde{q}_a^*\} \nn \\
&&\qquad\quad
+(X_1-iX_2)q_a^* \tilde{q}_a^* 
-\bar{\psi}_{q_a}(\phi^* -m_a^* )\bar{\psi}_{\tilde{q}_a}
-i\sqrt{2}\bar{\psi}_{\tilde{q}_a}\bar{\psi}q_a^* 
-i\sqrt{2}\bar{\psi}_{q_a}\bar{\psi}
\tilde{q}_a^* \Bigr] 
%\nn \\&&
\, . \label{lag4}
\end{eqnarray}
This is the full Lagrangian including fermion terms compared to 
the bosonic one in Eq.(\ref{lag2}). 

To obtain the ${\cal N}=1$ SUSY transformations in the 
Wess-Zumino gauge, 
one has to combine 
an ordinary SUSY transformation 
with an accompanying 
gauge transformation to preserve the Wess-Zumino gauge. 
Let us consider the four SUSY transformations 
 $\delta_{\xi_+}$ given by 
 an infinitesimal Grassmann number 
$\xi_+$ in the dierction of 
$\theta_+$ in Eqs.(\ref{vector1})--(\ref{hyper2}). 
The ${\cal N}=2$ vector multiplet represented by 
superfields (\ref{vector1}) and (\ref{vector2}) 
transforms under 
the infinitesimal SUSY $\delta_{\xi_+}$ along 
the left-handed spinor $\theta_+$ in the 
Wess-Zumino gauge as \cite{WB} 
\begin{eqnarray}
\delta_{\xi_+} v^m &=& 
i\bar{\xi}_+\bar{\sigma}^m\lambda+i\xi_+\sigma^m\bar{\lambda} , 
\label{SUSYvL1}\\
\delta_{\xi_+} \lambda &=& 
\sigma^{mn}v_{mn}\xi_++iX_3\xi_+ , 
\label{SUSYvL2} \\
\delta_{\xi_+}X_3&=&
\bar{\xi}_+\bar{\sigma}^m\partial_m\lambda 
-\xi_+\sigma^m\partial_m\bar{\lambda} , 
\label{SUSYvL3} \\
\delta_{\xi_+}\phi &=&
\sqrt{2}\xi_+(-i\sqrt{2}\psi) , 
\label{SUSYvL4} \\
\delta_{\xi_+}(-i\sqrt{2}\psi )&=&
i\sqrt{2}\sigma^m\partial_m \phi \bar{\xi}_+
+\sqrt{2}(X_1+iX_2)\xi_+ , 
\label{SUSYvL5} \\
\delta_{\xi_+}(X_1+iX_2)&=&
i\sqrt{2}\bar{\xi}_+\bar{\sigma}^m\partial_m 
(-i\sqrt{2}\psi ) \, . 
\label{SUSYvL6}
\end{eqnarray}
Similarly we obtain the supersymmetry transformation 
rules for 
hypermultiplets in the Wess-Zumino gauge 
\begin{eqnarray}
\delta_{\xi_+}q_a&=&\sqrt{2}\xi_+\psi_{q_a} , 
\label{SUSYhL1}\\
\delta_{\xi_+}\psi_{q_a}&=&
i\sqrt{2}\sigma^m{\cal D}_m q_a\bar{\xi}_+
+\sqrt{2}(F^\prime_a-\left(\phi^*-m_a^*\right) \tilde{q}^*_a )\xi_+ , 
\label{SUSYhL2}\\
\delta_{\xi_+}(F^\prime_a-\left(\phi^*-m_a^*\right) \tilde{q}^*_a)&=&
i\sqrt{2}\bar{\xi}_+\bar{\sigma}^m{\cal D}_m\psi_{q_a}
+2i\bar{\xi}_+\bar{\lambda}q_a , 
\label{SUSYhL3}\\
\delta_{\xi_+}\tilde{q}_a&=&
\sqrt{2}\xi_+\psi_{\tilde{q}_a} 
%\delta_{\xi_+}\tilde{q}^*_a&=&
%\sqrt{2}\bar{\xi}_+\bar{\psi}_{\tilde{q}_a} 
, \label{SUSYhL4}\\
\delta_{\xi_+}\psi_{\tilde{q}_a}&=&
i\sqrt{2}\sigma^m{\cal D}_m\tilde{q}_a\bar{\xi}_+
+\sqrt{2}(-\tilde{F}^{\prime}_a
-\left(\phi^*-m_a^*\right) q^*_a)\xi_+ 
%\delta_{\xi_+}\bar{\psi}_{\tilde{q}_a}&=&
%i\sqrt{2}\bar{\sigma}^m{\cal D}_m\tilde{q}^*_a\xi_+
%+\sqrt{2}(-\tilde{F}^{\prime*}_a-\left(\phi-m_a\right) q_a)\bar{\xi}_+ 
, 
\label{SUSYhL5}\\
\delta_{\xi_+}(-\tilde{F}^{\prime}_a-\left(\phi^*-m_a^*\right) q^*_a)&=&
i\sqrt{2}\bar{\xi}_+\bar{\sigma}^m
{\cal D}_m\psi_{\tilde{q}_a}
-2i\bar{\xi}_+\bar{\lambda}\tilde{q}_a 
%\delta_{\xi_+}(-\tilde{F}^{\prime*}_a-\left(\phi-m_a\right) q_a)&=&
%i\sqrt{2}\xi_+\sigma^m{\cal D}_m\bar{\psi}_{\tilde{q}_a}
%+2i\xi_+\lambda\tilde{q}^*_a 
\, , \label{SUSYhL6}
\end{eqnarray}

Let us now consider  the remaining  
four SUSY transformations $\delta_{\xi_-}$ 
also along the left-handed spinor $\theta_-$ 
which are not manifest by the 
above ${\cal N}=1$ superfield formalism 
(\ref{lag1}) or the associated component 
formalism (\ref{lag4}) 
in the Wess-Zumino gauge. 
%The corresponding Grassmann number 
%is denoted as $\theta_-$. 
Please note that Grassmann numbers $\xi_-, \theta_-$ are 
left-handed chiral spinors, similarly to $\xi_+, \theta_+$ 
and not to be confused with the right-handed 
spinors. %, in spite of the suffix $-$. 
To work out  $\delta_{\xi_-}$, 
we shall follow the classical method of Fayet \cite{Fayet}. 
The two sets of left-handed chiral spinor Grassmann numbers 
%supertransformations  $\delta_{\xi_+}$ and  $\delta_{\xi_-}$ 
$\theta_+$ and $\theta_-$ 
should form a doublet under 
an internal $SU(2)_R$ transformation 
$\mathfrak{M}$ 
whose representation matrix is denoted by 
 a $2 \times 2$ matrix $M$
\begin{eqnarray}
\mathfrak{M}\left(
\begin{array}{c}
\theta_+ \\
\theta_-
%Q_+ \\Q_-
\end{array}
\right)\mathfrak{M}^{-1}=
M\left(
\begin{array}{c}
\theta_+ \\
\theta_-
%Q_+ \\Q_-
\end{array}
\right) \, . \label{sym1}
\end{eqnarray}
It is enough to consider a discrete transformation, for instance, 
an $SU(2)_R$ rotation around second axis by $\pi$ 
\begin{eqnarray}
\mathfrak{M}_0=\exp (i\pi I_2) , 
\qquad 
M_0=i\sigma_2=\left(
\begin{array}{cc}
0 & 1 \\
-1 & 0
\end{array}
\right) \, . \label{eq:RParity}
\end{eqnarray}
Following Fayet, we demand that
$v_m$, $\phi$, $\psi_{q_a}$, 
and $\psi_{\tilde{q}_a}$ be $SU(2)_R$ singlets, and that 
\begin{eqnarray}
\left(
\begin{array}{c}
\lambda
\\
%i
\psi 
\end{array}
\right) \, ,
\qquad 
\left(
\begin{array}{c}
q_a \\
%i
\tilde{q}_a^*
\end{array}
\right) \, ,
\qquad 
\left(
\begin{array}{c}
-%i
\tilde{q}_a \\
%-
q^*_a
\end{array}
\right) \, , \label{sym4}
\end{eqnarray}
be $SU(2)_R$ doublets 
. 
The equations of motion (\ref{eom1}) and (\ref{eom2}) 
show that
\begin{eqnarray}
\left(
\begin{array}{c}
X_1 \\
-X_2 \\
%-X_2 \\-X_1 \\
X_3
\end{array}
\right) \label{sym6}
\end{eqnarray}
transforms as an $SU(2)_R$ triplet. 
The equations of motion for auxiliary fields in 
Eqs.(\ref{eom3}), (\ref{eom4}) for the 
hypermultiplets become 
\begin{equation}
F^{'*}_a 
=F^{*}_a+\left(\phi-m_a\right)\tilde q_a =
0 
%m_a\tilde q_a
, 
\qquad 
\tilde F^{'}_a 
=-\tilde F_a - q^*_a\left(\phi^*-m_a^*\right) 
=0 %-m^*_a q^*_a
, 
\end{equation}
which suggest that 
the auxiliary fields in hypermultiplets generate 
$SU(2)_R$ doublets
\begin{eqnarray}
\left(
\begin{array}{c}
\left(m_a^*-\phi^*\right) q_a
-
\tilde{F}^{\prime*}_a 
\\
\left(m_a^*-\phi^*\right)\tilde{q}^*_a 
+
F^\prime_a
\end{array}
\right)
\sim 
\left(
\begin{array}{c}
 q_a
\\
\tilde{q}^*_a 
\end{array}
\right)
\, ,
\qquad 
\left(
\begin{array}{c}
-\tilde{q}_a\left(m_a^*-\phi^*\right)
-F^{\prime*}_a 
\\
q^*_a\left(m_a^*-\phi^*\right)
-\tilde{F}^\prime_a
\end{array}
\right)
\sim 
\left(
\begin{array}{c}
-\tilde{q}_a
\\
q^*_a 
\end{array}
\right)
. \label{sym7}
\end{eqnarray}

\begin{figure}[htbp]
\begin{center}
	\scalebox{1.0}{\includegraphics{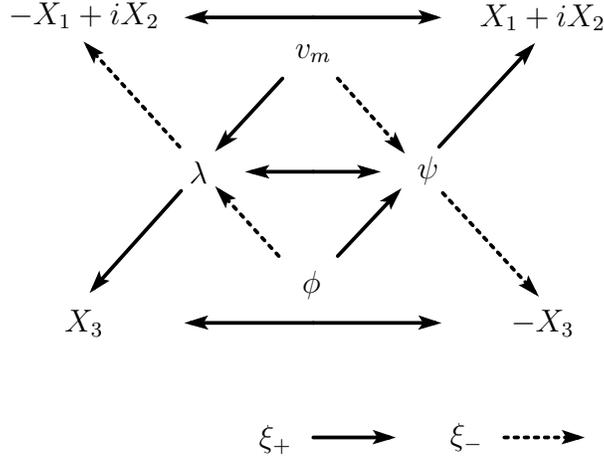}}
	\caption{The transformation laws for vector multiplet}
	\label{VM}
\end{center}
\end{figure}

\begin{figure}[htbp]
\begin{center}
	\scalebox{0.9}{\includegraphics{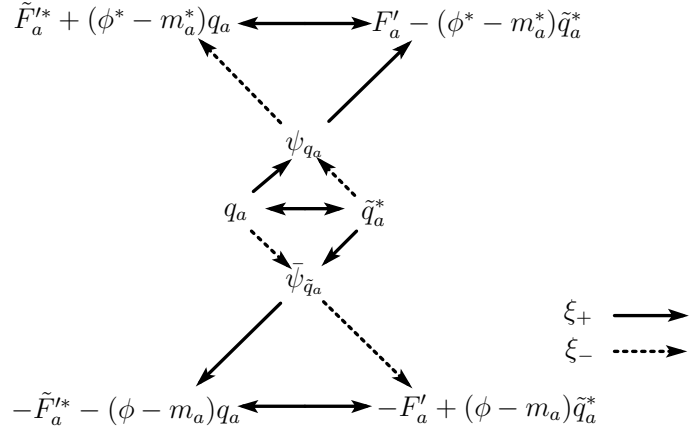}}
%	\scalebox{1.0}{\includegraphics{./hyper.eps}}
	\caption{The transformation laws for hypermultiplet}
	\label{HM}
\end{center}
\end{figure}

By applying the discrete $SU(2)_R$ internal transformation 
(\ref{eq:RParity}) with the assignment (\ref{sym4})--(\ref{sym7}) 
to the first four SUSY (\ref{SUSYvL1})--(\ref{SUSYhL6}) 
represented by an infinitesimal 
$\xi_+$, 
we can find the another four SUSY  $\delta_{\xi_-}$ 
corresponding to an infinitesimal 
Grassmann number $\xi_-$. 
 We obtain  $\delta_{\xi_-}$ transformations of the 
 ${\cal N}=2$ vector multiplet 
in the Wess-Zumino gauge 
\begin{eqnarray}
\delta_{\xi_-} v^m &=&
i\xi_-\sigma^m
\bar{\psi}
+i\bar{\xi}_-\bar{\sigma}^m
\psi
\label{SUSYvR1}\\
\delta_{\xi_-}
\psi
 &=&
\sigma^{mn}v_{mn}\xi_--iX_3\xi_- , \label{SUSYvR2}\\
\delta_{\xi_-}(-X_3)&=&
-\xi_-\sigma^m\partial_m
\bar{\psi}
+\bar{\xi}_-\bar{\sigma}^m\partial_m
\psi
\label{SUSYvR3}\\
\delta_{\xi_-}\phi &=&
\sqrt{2}\xi_-(\sqrt{2}i\lambda) , \label{SUSYvR4}\\
\delta_{\xi_-}(\sqrt{2}i\lambda )&=&
i\sqrt{2}\sigma^m\partial_m \phi\bar{\xi}_-
+\sqrt{2}(-X_1+iX_2)\xi_- , 
\label{SUSYvR5}\\ 
\delta_{\xi_-}(-X_1+iX_2)&=&
i\sqrt{2}\bar{\xi}_-\bar{\sigma}^m
\partial_m(\sqrt{2}i\lambda )\, . 
\label{SUSYvR6}
\end{eqnarray}
The transformation laws for the ${\cal N}=2$ 
vector multiplet 
(\ref{SUSYvL1})--(\ref{SUSYvL6}) with respect 
to the Grassmann numbers 
$\theta_+$ and 
(\ref{SUSYvR1})--(\ref{SUSYvR6}) 
with respect to the Grassmann numbers 
$\theta_-$ are illustrated in Fig.\ref{VM}. 
 For hypermultiplet,  $\delta_{\xi_-}$ transformations in 
 the Wess-Zumino gauge 
is given by 
\begin{eqnarray}
\delta_{\xi_-}\tilde{q}^*_a&=&
\sqrt{2}\xi_-
%(-i
\psi_{q_a}
, \label{SUSYhR4}\\
\delta_{\xi_-}
\psi_{q_a}
&=&
i\sqrt{2}\sigma^m{\cal D}_m \tilde{q}^*_a\bar{\xi}_-
+\sqrt{2}\left(\tilde{F}^{\prime*}_a
+\left(\phi^*-m_a^*\right) q_a\right)\xi_- , 
\label{SUSYhR5}\\
\delta_{\xi_-}\left(\tilde{F}^{\prime*}_a
+\left(\phi^*-m_a^*\right) q_a\right)&=&
i\sqrt{2}\bar{\xi}_-\bar{\sigma}^m{\cal D}_m
\psi_{q_a}
+2i\bar{\xi}_-
\bar{\psi}
\tilde{q}^*_a, \label{SUSYhR6}\\
\delta_{\xi_-}q^*_a&=&
\sqrt{2}\xi_-(-\psi_{\tilde{q}_a}) , 
\label{SUSYhR1}\\
\delta_{\xi_-}(-\psi_{\tilde{q}_a})&=&
i\sqrt{2}\sigma^m{\cal D}_m q^*_a\bar{\xi}_-
+\sqrt{2}\left(-F^{\prime*}_a
+ \left(\phi^*-m_a^*\right) \tilde{q}_a\right)\xi_- , 
\label{SUSYhR2}\\
\delta_{\xi_-}\left(-F^{\prime*}_a
+ \left(\phi^*-m_a^*\right) \tilde{q}_a\right)&=&
i\sqrt{2}\bar{\xi}_-\bar{\sigma}^m
{\cal D}_m(-\psi_{\tilde{q}_a})
-2i\bar{\xi}_-\bar{\psi} q^*_a  \, . 
\label{SUSYhR3}
\end{eqnarray}
%One should note that the factor $\frac{m_a^*}{m_a}$ 
%multiplying the auxiliary fields appears 
%in Eqs.(\ref{SUSYhR3}) and (\ref{SUSYhR6}). 
%We shall show in sect.\ref{sc:DimRed} 
%that this factor can also be understood by considering 
%the dimensional reduction from five dimensions. 
The transformation law for the ${\cal N}=2$ hypermultiplet 
(\ref{SUSYhL1})--(\ref{SUSYhL6}) with respect to the Grassmann numbers 
$\theta_+$ and 
(\ref{SUSYhR1})--(\ref{SUSYhR6}) 
with respect to the Grassmann numbers 
$\theta_-$ are illustrated in Fig.\ref{HM}. 

To summarize the transformation property (\ref{SUSYvR1})--(\ref{SUSYhR6}) 
under  $\delta_{\xi_-}$, it is convenient 
to define the 
 following superfields using another set of 
Grassmann number 
$\theta_-$ 
\begin{eqnarray}
V_-&=&
-\theta_-\sigma^m \bar{\theta}_-v_m 
+i\theta_-^2\bar{\theta}_-
%(-i
\bar{\psi}
%) 
-i\bar{\theta}_-^2\theta_-
%(i
\psi
%) 
+\frac{1}{2}\theta_-^2\bar{\theta}_-^2(-X_3) , 
\label{vector3}\\
\Phi_-&=&
\phi +\sqrt{2}\theta_-(\sqrt{2}i\lambda )
+\theta_-^2(-X_1+iX_2) , 
\label{vector4}\\
Q_{-a}&=&
\tilde{q}_a^* +\sqrt{2}\theta_-
%(-i
\psi_{q_a}
%)
+\theta_-^2(
%-
\tilde{F}^{\prime*}_a
%-
+\left(\phi^*-m_a^*\right) q_a ) , 
\label{hyper3}\\
\tilde{Q}_{-a}&=&
q^*_a+\sqrt{2}\theta_-(
%i
-\psi_{\tilde{q}_a})
+\theta_-^2(-F^{\prime*}_a
+\tilde{q}_a \left(\phi^*-m_a^*\right) )\, . 
\label{hyper4}
\end{eqnarray}

We can now rewrite 
the Lagrangian (\ref{lag4}) to make the second set of 
four SUSY transformations $\delta_{\xi_-}$ 
(\ref{SUSYvR1})--(\ref{SUSYhR6}) 
manifest 
\begin{eqnarray}
{\cal L}&=&
-\frac{1}{4g^2}v_{mn}v^{mn}
-\frac{1}{g^2}i\bar{\psi}\bar{\sigma}^m\partial_m\psi 
+\frac{1}{2g^2}(-X_3)^2 +\frac{1}{2g^2}|X_1-iX_2|^2 
\nn \\
&&
-\frac{1}{2g^2}|\partial_m \phi |^2 
-\frac{1}{g^2}i\lambda\sigma^m\partial_m\bar{\lambda}  
+c(-X_3)+b (-X_1-iX_2)+b^* (-X_1+iX_2) 
\nn \\
&&+\sum_{a=1}^n\Bigl[(-\tilde{F}^\prime_a-q^*_a\left(\phi-m_a\right))
(-\tilde{F}_a^{\prime*} -\left(\phi^*-m_a^*\right) q_a)
-|{\cal D}_m \tilde{q}_a|^2  
\nn \\
&&\qquad\quad-i\bar{\psi}_{q_a}\bar{\sigma}^m
{\cal D}_m\psi_{q_a}
-i\sqrt{2}(\bar{\psi}_{q_a}\bar{\psi}\tilde{q}_a^* 
-\psi_{q_a}\psi\tilde{q}_a) +(-X_3)|\tilde{q}_a|^2\Bigr]  
\nn \\
&&+\sum_{a=1}^n\Bigl[(F_a^{\prime*} 
-\tilde{q}_a\left(\phi^*-m_a^*\right))(
F^\prime_a-\left(\phi-m_a\right)\tilde{q}_a^*)
-|{\cal D}_m q_a|^2  
\nn \\
&&\qquad\quad
-i\psi_{\tilde{q}_a}\sigma^m{\cal D}_m
\bar{\psi}_{\tilde{q}_a}
+i\sqrt{2}(-\bar{\psi}_{\tilde{q}_a}\bar{\psi}q_a^* 
+\psi_{\tilde{q}_a}\psi q_a) -(-X_3)|q_a|^2 \Bigr]  
\nn \\
&&+\sum_{a=1}^n\Bigl[(\phi -m_a)\{q_a^* 
(-\tilde{F}^{\prime*}_a
-\left(\phi^*-m_a^*\right) q_a)+ (F_a^{\prime*} 
-\tilde{q}_a\left(\phi^*-m_a^*\right) )\tilde{q}_a^*\}  
\nn \\
&&\qquad\quad +(X_1-iX_2)q_a^*\tilde{q}_a^* 
-\psi_{\tilde{q}_a}(\phi -m_a)\psi_{q_a}
-i\sqrt{2}\psi_{\tilde{q}_a}\lambda\tilde{q}_a^* 
+i\sqrt{2}\psi_{q_a}\lambda q_a^* \Bigr]  
\nn \\
&&+\sum_{a=1}^n\Bigl[(\phi^* -m^*_a)\{\tilde{q}_a
(F_a-\left(\phi-m_a\right)\tilde{q}^* )
+(-\tilde{F}^\prime_a -q^*_a\left(\phi-m_a\right) )q_a\}  
\nn \\
&&\qquad\quad 
+(X_1+iX_2)\tilde{q}_aq_a 
-\bar{\psi}_{q_a}(\phi^* -m_a^* )\bar{\psi}_{\tilde{q}_a}
+i\sqrt{2}\bar{\psi}_{\tilde{q}_a}\bar{\lambda}\tilde{q}_a
-i\sqrt{2}\bar{\psi}_{q_a}\bar{\lambda}q_a \Bigr]  
%\nn \\&&
 \, . 
\label{lag5}
\end{eqnarray}
We can finally assemble the above component form of 
the Lagrangian 
to an ${\cal N}=1$ superfield formalism using the 
second set of 
Grassmann number $\theta_-$ in 
Eqs.(\ref{vector3})--(\ref{hyper4}), 
in contrast to those superfields with $\theta_+$ in 
Eqs.(\ref{vector1})--(\ref{hyper2}) 
\begin{eqnarray}
{\cal L}
&\!\!\!\!=&\!\!\!\!
\frac{1}{4g^2}\Bigl(W_-^\alpha W^-_\alpha 
\Big|_{\theta_-^2}+\bar{W}^-_{\dot \alpha}
\bar{W}_-^{\dot \alpha}
\Big|_{\bar{\theta}_-^2}\Bigr)
+\frac{1}{2g^2}\Phi_-^\dagger\Phi_- 
\Big|_{\theta_-^2\bar{\theta}_-^2}
+\sum_{a=1}^n\Bigl(Q^\dagger_{-a}{\rm e}^{2V_-}Q_{-a}
+\tilde{Q}^\dagger_{-a}{\rm e}^{-2V_-}\tilde{Q}_{-a}\Bigr)
\Big|_{\theta_-^2\bar{\theta}_-^2} 
\nonumber \\
&\!\!\!\! &\!\!\!\!
+2c V_- \Big|_{\theta_-^2\bar{\theta}_-^2}
+\left(\sum_{a=1}^n\Bigl(\Phi_- -m_a\Bigr)
Q_{-a}\tilde{Q}_{-a}
\Big|_{\theta_-^2}
%\nonumber \\ \!\!& &\!\!
+b^*\,\Phi_- \Big|_{\theta_-^2}
+\textrm{h.c.}\right)
 \, . \label{eq:lag_R}
\end{eqnarray}

Let us note that the Lagrangian is not invariant under 
the automorphism $SU(2)_R$ if FI parameters are 
present. 
Obviously the Lagrangian is invariant under the eight 
SUSY transformations when it is invariant under 
the first set of four SUSY and the discrete $SU(2)_R$ 
transformation ${\frak M}_0$. 
However, it is important to realize that 
the Lagrangian represented by 
(\ref{lag4}) and (\ref{eq:lag_R}) 
is invariant under all the eight SUSY transformations 
(\ref{SUSYvL1})--(\ref{SUSYhL6}) and 
(\ref{SUSYvR1})--(\ref{SUSYhR6}) 
irrespective of the values of the FI parameters 
$c$ and $b$. 
This is because the difference of the Lagrangian 
transformed by ${\frak M}_0$ and the original one 
is given by the FI terms which are transformed to 
total derivative by SUSY transformations 
 $\delta_{\xi_+}$ and  $\delta_{\xi_-}$. 
Therefore the action is invariant as usual for the 
SUSY theories.

%======================================================
%	Section [2 SUSY out of 8 SUSY conserved (temporary)]
%	* 2 SUSY out of 8 SUSY conserved
%	* 
%	* 
%======================================================
%%%%%%%%%%%%%%%%%%%%%%%%%%%%%%%%%%%%%%%%%%%%%%%%%%%%%%%%%
%\section{2 SUSY out of 8 SUSY conserved}

%%%%%%%%%%%%%%%%%%%%%%%%%%%%%%%%%%%%%%%%%%%%%%%%%%%%%%%%%

Theories with eight SUSY like our model 
have been shown to possess three complex structures 
\cite{AlvarezFreedman}, \cite{GPT}. 
Our formulation in terms of two sets of Grassmann numbers 
$\theta_+, \theta_-$ does not make this property manifest. 
In fact our choice of $c=0$, $b \in {\bf R}$ 
breaks $SU(2)_R$ symmetry and particular complex structure 
has been selected. 
However, this particular choice of complex structure will 
turn out to be useful for the analysis of our model and 
solution. 
For instance, we will show that we can choose 
one of the two conserved SUSY directions from 
$\theta_+$, and the other from $\theta_-$. 

Since all the eight SUSY transformations are clarified, 
we are now in a position to determine precisely how 
many SUSY charges out of these eight are conserved 
by our solution (\ref{sol11})--(\ref{sol12}) 
of the domain wall junction. 
We have already found in Eq.(\ref{eq:quarter1}) 
that one out of  four 
SUSY  $\delta_{\xi_-}$ is conserved. 
We need to examine another four SUSY  $\delta_{\xi_-}$. 
Let us first consider fermions ${\psi}_{\tilde{q}_a}$ 
and $\psi_{q_a}$ 
in the ${\cal N}=1$ 
chiral scalar superfields $\tilde{Q}_{-a}$ in Eq.(\ref{hyper3}) 
and $Q_{-a}$ in Eq.(\ref{hyper4}) for the hypermultiplets 
\begin{eqnarray}
\delta_{\xi_-}
%(-i
\psi_{q_a}
%)
&=&
i\sqrt{2}\sigma^m\partial_m \tilde{q}^*_a\bar{\xi}_-
+\sqrt{2}(\phi^* - m_a^*) q_a\xi_- 
\nn\\
&=&i2\sqrt{2}\Bigl(\sigma^{+}\partial_{z}
\tilde{q}^*_a\bar{\xi}_-
+\sigma^{-}\partial_{\bar{z}}\tilde{q}^*_a\bar{\xi}_-
-i\frac{1}{2}(\phi^* -m_a^* ) q_a\xi_-\Bigr) 
\nn \\
&=&i2\sqrt{2}(\sigma^{+}\bar{\xi}_- +i\xi_-)
\partial_{z}\tilde{q}^*_a
+i2\sqrt{2}\sigma^{-}\bar{\xi}_-\partial_{\bar{z}}
\tilde{q}^*_a \, ,
\end{eqnarray}
\begin{eqnarray}
\delta_{\xi_-}(-\psi_{\tilde{q}_a})&=&
i\sqrt{2}\sigma^m\partial_m q^*_a\bar{\xi}_-
+\sqrt{2}(\phi^*-m^*_a)\tilde{q}_a )\xi_- 
\nn \\
&=&i2\sqrt{2}\Bigl(\sigma^{+}\partial_{z} q^*_a\bar{\xi}_-
+\sigma^{-}\partial_{\bar{z}} q^*_a\bar{\xi}_-
-i\frac{1}{2}(\phi^* -m^*_a)\tilde{q}_a\xi_-\Bigr)
\nn \\
&=&i2\sqrt{2}\sigma^{-}\bar{\xi}_-\partial_{\bar{z}} q^*_a
+i2\sqrt{2}(\sigma^{+}\bar{\xi}_- +i\xi_-)
\partial_{z} q^*_a .
\end{eqnarray}
Therefore the conserved SUSY direction is given by 
\begin{eqnarray}
&& \sigma^{+}\bar{\xi}_- = 
-i\xi_- \quad \textrm{and}\quad \sigma^{-}\bar{\xi}_-=0 \, .
\label{eq:SUSY_R}
\end{eqnarray}

We have to study the SUSY transformation of 
the remaining fermions 
$\lambda$ and $\bar\psi$ in the ${\cal N}=1$ 
chiral scalar superfield $\Phi_-$ and $V_-$ 
using 
the BPS equations  
(\ref{PSC5}) and (\ref{eq:PSC_N1}) with $\Omega=-1$ and 
algebraic equations of motion for auxiliary fields 
(\ref{eq:aux_N1}). 
Moreover, we note that 
%the ${\frak M}_0$ odd 
%auxiliary fields vanish 
$X_2=X_3=0$ in our solution. 
Then we find the same SUSY directions are conserved for 
$\lambda$ 
\begin{eqnarray}
\delta_{\xi_-}(i\sqrt{2}\lambda)&=&
i\sqrt{2}\sigma^m\partial_m\phi\bar{\xi}_-
+\sqrt{2}(-X_1+iX_2)\xi_- , 
\nn \\
&=&
i2\sqrt{2}(\sigma^{+}\partial_{z}
+\sigma^{-}\partial_{\bar{z}})\phi\bar{\xi}_-
+\sqrt{2}(-X_1+iX_2)\xi_- , 
\nn \\
&=&
i2\sqrt{2}(\sigma^{+}\bar{\xi}_-+i\xi_-)\partial_{z}\phi 
+i2\sqrt{2}\sigma^{-}\bar{\xi}_-\partial_{\bar{z}}\phi =0 ,
\end{eqnarray}
with the same infinitesimal SUSY transformation parameters 
(\ref{eq:SUSY_R}). 
The right-hand side of $\xi_-$ transformation of $\psi$ 
in Eq.(\ref{SUSYvR2}) involve only 
%components of the ${\cal N}=1$ vector multiplet 
$v_m$ and $X_3$ which vanish in our solution. 
Therefore they conserve all the SUSY transformations 
trivially. 
%\footnote{In order to preserve the same SUSY direction as in 
%Eq.(\ref{eq:SUSY_R}), it seems necessary to have $X_3=0$ 
%as in our solution of domain wall junction 
%(\ref{eq:vector_D_cond1}). }. 

Summarizing our results for all the eight SUSY, we see that 
there are two conserved directions in the 
Grassmann parameter
\begin{eqnarray}
&& \sigma^{+}\bar{\xi}_+ = 
i\xi_+ \quad \textrm{and}\quad \sigma^{-}\bar{\xi}_+=0 , \\ 
&& \sigma^{+}\bar{\xi}_- = 
-i\xi_- \quad \textrm{and}\quad \sigma^{-}\bar{\xi}_-=0 \, .
\end{eqnarray}
Namely we have determined the two conserved directions
\begin{eqnarray}
\xi_{+1}=i(\xi_{+1})^* \, ,\quad 
\xi_{+2}=0 \, ,\quad 
\xi_{-1}=-i(\xi_{-1})^* \, ,
\quad \xi_{-2}=0
\end{eqnarray}
where we set
\begin{eqnarray}
&&\xi_+=\left(
\begin{array}{c}
\xi_{+1} \\
\xi_{+2}
\end{array}
\right)\, ,
\qquad 
\bar{\xi}_{+}=\left(
\begin{array}{c}
\bar{\xi}_{+}^{\dot{1}} \\
\bar{\xi}_{+}^{\dot{2}}
\end{array}
\right)=\left(
\begin{array}{c}
(\xi_{+2})^* \\
-(\xi_{+1})^*
\end{array}
\right)\, , \\
&&\xi_-=\left(
\begin{array}{c}
\xi_{-1} \\
\xi_{-2}
\end{array}
\right)\, ,
\qquad 
\bar{\xi}_{-}=\left(
\begin{array}{c}
\bar{\xi}_{-}^{\dot{1}} \\
\bar{\xi}_{-}^{\dot{2}}
\end{array}
\right)=\left(
\begin{array}{c}
(\xi_{-2})^* \\
-(\xi_{-1})^*
\end{array}
\right) \, .
\label{eq:20SUSY}
\end{eqnarray}
We have now established that our domain wall solution 
preserves two out of eight SUSY. 
%Moreover, it is interesting to note that the conserved SUSY 
%is $(2, 0)$ SUSY instead of $(1, 1)$ as one can recognize in 
%Eq.(\ref{eq:20SUSY}) that the two conserved SUSY charges 
%have the same chirality. 
%Therefore the effective low-energy theory on the background 
%of our domain wall junction is chiral, similarly to the 
%${\cal N}=1$ SUSY case \cite{OINS}, \cite{INOS}. 

%%%%%%%%%%%%%%%%%%%%%%%%%%%%%%%%%%%%%%%%%%%%%%%%%%%%%%%%%
\section{Dimensional reduction from five dimensions
}
\label{sc:DimRed}

%%%%%%%%%%%%%%%%%%%%%%%%%%%%%%%%%%%%%%%%%%%%%%%%%%%%%%%%%

The ${\cal N}=1$ superfield formalism is of course most 
useful to 
study ${\cal N}=1$ SUSY theories in four dimensions. 
It can also be used to describe the ${\cal N}=2$ SUSY 
theories in four dimensions making only four out of 
eight SUSY manifest. 
In order to describe ${\cal N}=2$ SUSY theories in 
five dimensions, however, we need to sacrifice the 
five-dimensional Lorentz invariance \cite{MirabelliPeskin}, 
\cite{AGW}--\cite{LLP}. 
In terms of components, we can always express the ${\cal N}=2$ 
SUSY theories with all the necessary auxiliary fields to close 
the algebra off-shell in the Wess-Zumino gauge. 

The highest dimension allowed by the ${\cal N}=2$ SUSY 
is six. 
The hypermultiplet in six dimensions cannot have masses. 
The five-dimensional theories can be obtained by 
a dimensional reduction from six dimensions. 
If we perform a nontrivial dimensional reduction, allowing 
the momenta in the sixth dimension, we obtain a massive 
hypermultiplet \cite{SierraTownsend}. 
Therefore the hypermultiplets in five dimensions 
can have only real mass parameters. 
The ${\cal N}=2$ SUSY theories in four dimensions 
can be obtained by dimensionally reducing the five-dimensional 
theories. 
The real scalar field $\Sigma$ in five dimensions 
comes originally from the 
sixth component of gauge field and 
the combination 
$\Sigma +iv_5$ becomes a complex scalar in four dimensions, 
when we consider 
the reduction to four dimensions.
To obtain the four-dimensional theory with complex 
mass parameters, 
we should perform a nontrivial dimensional reduction in 
the $x^5$ direction. 
In this process, the mass terms arise as momenta 
in the fifth dimension. 
Conversely we can recover the five-dimensional theory 
by restoring the fifth dimension from the imaginary 
part of the complex mass parameter 
$m_a\equiv m_{aR}+im_{aI}$ as 
\begin{equation}
\partial_5 q_{a}=-im_{aI} q_{a}, 
\qquad 
\partial_5 \psi_{q_{a}}=-im_{aI} \psi_{q_{a}}. 
\end{equation}
The imaginary part of the complex scalar field $\phi$ 
can also be identified as the fifth component of the 
vector field $v_5$ as 
\begin{equation}
\phi=\Sigma + i v_5. 
\end{equation}
We can recover the covariant derivative along the fifth 
direction as 
\begin{equation}
i(v_5-m_{aI}) q_{a} 
=
(\partial_5+iv_5) q_{a}
=
 {\cal D}_5 q_{a}
\end{equation}
Therefore the 
mass terms associated with the quark 
fields are reduced to covariant derivatives 
in the fifth dimension 
\begin{equation}
(\phi-m_a) q_{a}
=
\left[(\Sigma - m_{aR})+i(v_5-m_{aI})\right] q_{a} 
=
\left[(\Sigma - m_{aR})+ {\cal D}_5 \right] q_{a}
\end{equation}

In the spirit of the ${\cal N}=1$ superfield formalisms, 
we can express the 
${\cal N}=2$ vector and hypermultiplets in five dimensions 
by two kinds of superfields similarly to our results 
(\ref{vector1})--(\ref{hyper2}), and 
(\ref{vector3})--(\ref{hyper4}) 
for four-dimensional ${\cal N}=2$ theories. 
The $\theta_+$ superfields are given by 
\begin{eqnarray}
V_+(x,\theta_+, \bar \theta_+)
&=&
-\theta_+\sigma^m\bar{\theta}_+v_m
+i\theta_+^2\bar{\theta}_+\bar{\lambda}
-i\bar{\theta}_+^2\theta_+\lambda
+\frac{1}{2}\theta_+^2\bar{\theta}_+^2
(X_3-\partial_5\Sigma) , 
\label{eq:5dSYM-SFHL1}
\\
\Phi_+(y,\theta_+ )
&=&
(\Sigma +iv_5)
+\sqrt{2}\theta_+(-i\sqrt{2}\psi)
+\theta_+^2(X_1+iX_2) , 
\label{eq:5dSYM-SFHL2}
\\
Q_{+a}(y,\theta_+ )
&=&
q_{a}+\sqrt{2}\theta_+\psi_{q_{a}}
+\theta_+^2(F^\prime_{a} +{\cal D}_5\tilde{q}^\ast_{a} 
-(\Sigma-m_{aR}) 
\,\tilde{q}^\ast) , 
\label{eq:5dSYM-SFHL3}
\\
\tilde{Q}_{+a}(y,\theta_+ )
&=&
\tilde{q}_{a} +\sqrt{2}\theta_+\psi_{\tilde{q}_{a}}
+\theta_+^2(-\tilde{F}^{\prime}_{a}
-{\cal D}_5 q^\ast_{a} -q^\ast_{a}(\Sigma-m_{aR}) ) . 
\label{eq:5dSYM-SFHL4}
\end{eqnarray}
The $\theta_-$ superfields are given by 
\begin{eqnarray}
V_-(x,\theta_-, \bar \theta_-)
&=&
-\theta_-\sigma^m\bar{\theta}_- v_m
+i\theta_-^2\bar{\theta}_-\bar{\psi}
-i\bar{\theta}_-^2\theta_- \psi 
+\frac{1}{2}\theta_-^2\bar{\theta}_-^2
(-X_3-\partial_5\Sigma) , 
\label{eq:5dSYM-SFHR1}
\\
\Phi_-(y,\theta_- )
&=&
(\Sigma +iv_5)
+\sqrt{2}\theta_-(\sqrt{2}i\lambda )
+\theta_-^2(-X_1+iX_2) , 
\label{eq:5dSYM-SFHR2}
\\
Q_{-a}(y,\theta_- )
&=&
\tilde{q}^\ast_{a} 
+\sqrt{2}\theta_-\psi_{q_{a}}
+\theta_-^2(\tilde{F}^{\prime\ast}_{a}-{\cal D}_5 q_{a} 
+(\Sigma-m_{aR}) \,q_{a} ) , 
\label{eq:5dSYM-SFHR3}
\\
\tilde{Q}_{-a} (y,\theta_- )
&=&
q^\ast_{a}
+\sqrt{2}\theta_-(-\psi_{\tilde{q}_{a}})
+\theta_-^2(-F^{\prime\ast}_{a}+{\cal D}_5 \tilde{q}_{a} 
+\tilde{q}\, (\Sigma-m_{aR}) ) \, .
\label{eq:5dSYM-SFHR4}
\end{eqnarray}
Please note that all the fields depend on the coordinate 
$x^5$ in fifth dimensions, in spite of almost the same 
appearance as the four-dimensional superfields. 

Since the mass term in five dimensions can be obtained as a nontrivial 
dimensional reduction from six dimensions, 
we can have only real mass parameter in the 
Lagrangian $m_{aR}$ \cite{Hebecker}, \cite{SierraTownsend}. 
Therefore we find the Lagrangian in terms of the $\theta_+$ 
${\cal N}=1$ 
superfields in Eqs.(\ref{eq:5dSYM-SFHL1})--
(\ref{eq:5dSYM-SFHL4}) as 
\begin{eqnarray}
\mathcal{L}
&\!\!\!\!=&\!\!\!\!
\frac{1}{4g^2}\Bigl(W_+^\alpha W^+_\alpha 
\Big|_{\theta_+^2}+\bar{W}^+_{\dot \alpha}\bar{W}_+^{\dot \alpha}
\Big|_{\bar{\theta}_+^2}\Bigr)
+\frac{1}{g^2}\left(\partial_5 V -{\Phi_+^\dagger+\Phi_+ \over 2} 
\right)\Big|_{\theta_+^2\bar{\theta}_+^2}
\nonumber \\
&\!\!\!\! &\!\!\!\!
+\sum_{a=1}^n\Bigl(Q^\dagger_{+a}{\rm e}^{2V_+}Q_{+a}
+\tilde{Q}^\dagger_{+a}{\rm e}^{-2V_+}\tilde{Q}_{+a}\Bigr)
\Big|_{\theta_+^2\bar{\theta}_+^2} 
-2c V_+ \Big|_{\theta_+^2\bar{\theta}_+^2}
\nonumber \\
&\!\!\!\! &\!\!\!\!
+\left(\sum_{a=1}^n\tilde{Q}_{+a}
\Bigl(\partial_5+\Phi_+ -m_a\Bigr)
Q_{+a}\Big|_{\theta_+^2}
-b\,\Phi_+ \Big|_{\theta_+^2}
+\textrm{h.c.}\right)
%\nonumber \\
 \, , \label{eq:lag-superfield+}
\end{eqnarray}
The same Lagrangian is given in terms of the $\theta_-$ 
${\cal N}=1$ superfields in Eqs.(\ref{eq:5dSYM-SFHR1})--
(\ref{eq:5dSYM-SFHR4}) as 
\begin{eqnarray}
\mathcal{L}
&\!\!\!\!=&\!\!\!\!
\frac{1}{4g^2}\Bigl(W_-^\alpha W^-_\alpha 
\Big|_{\theta_-^2}+\bar{W}^-_{\dot \alpha}\bar{W}_-^{\dot \alpha}
\Big|_{\bar{\theta}_-^2}\Bigr)
+\frac{1}{g^2}\left(\partial_5 V -{\Phi_-^\dagger+\Phi_- \over 2} 
\right)\Big|_{\theta_-^2\bar{\theta}_-^2}
\nonumber \\
&\!\!\!\! &\!\!\!\!
+\sum_{a=1}^n\Bigl(Q^\dagger_{+a}{\rm e}^{2V_-}Q_{+a}
+\tilde{Q}^\dagger_{+a}{\rm e}^{-2V_-}\tilde{Q}_{+a}\Bigr)
\Big|_{\theta_-^2\bar{\theta}_-^2} 
+2c V_- \Big|_{\theta_-^2\bar{\theta}_-^2}
\nonumber \\
&\!\!\!\! &\!\!\!\!
+\left(\sum_{a=1}^n\tilde{Q}_{+a}
\Bigl(\partial_5+\Phi_- -m_a\Bigr)
Q_{+a}\Big|_{\theta_-^2}
+b^*\,\Phi_- \Big|_{\theta_-^2}
+\textrm{h.c.}\right)
 \, , \label{eq:lag-superfield-}
\end{eqnarray}
In terms of components, we can express the Lagrangian 
more symmetrically with respect to the $SU(2)_R$ symmetry 
\begin{equation}
\mathcal{L}=
\mathcal{L}_{\rm boson}+
\mathcal{L}_{\rm fermion}
\label{5D-lagrangian}
\end{equation}
\begin{eqnarray}
\mathcal{L}_{\rm boson}&=&
-\frac{1}{4g^2}F_{MN}F^{MN}
-\frac{1}{2g^2}(\partial_M\Sigma)^2
+\frac{1}{2g^2}\{(X_1)^2+(X_2)^2+(X_3)^2\} 
\nn \\
&&
+c(-X_3)+b (-X_1-iX_2)+b^* (-X_1+iX_2) 
\nn \\
&&
+\sum_{a=1}^n \biggl[-\left|{\cal D}_M q_a\right|^2 
- \left|{\cal D}_M \tilde{q}_a\right|^2 
+ \left|F^\prime\right|^2
+ |\tilde{F}^{\prime}|^2
-(\Sigma - m_{aR})^2 (|q_a|^2+|\tilde q_a|^2)
\nn \\
&&+X_3(|q_a|^2 -|\tilde{q}_a|^2 )
+(X_1+iX_2)\tilde{q}_a q_a
+(X_1-iX_2)q^\ast_a \tilde{q}^\ast_a \biggr]
\label{5D-lag-boson}
\end{eqnarray}
\begin{eqnarray}
\mathcal{L}_{\rm fermion}
&
=
&
-\frac{1}{g^2}\lambda\bar{\sigma}^m\partial_m\bar{\lambda}
-\frac{1}{g^2}\bar{\psi}\bar{\sigma}^m\partial_m\psi 
-\frac{1}{g^2}\psi\partial_5\lambda
-\frac{1}{g^2}\bar{\psi}\partial_5\bar{\lambda}
 \nn \\
&&
+\sum_{a=1}^n\biggl[
-i\bar{\psi}_{q_a}\bar{\sigma}^m{\cal D}_m \psi_{q_a}
-i\psi_{\tilde{q}_a}\sigma^m {\cal D}_m\bar{\psi}_{\tilde{q}_a}
-\psi_{\tilde{q}_a} {\cal D}_5 \psi_{q_a}
-\bar{\psi}_{\tilde{q}_a} {\cal D}_5 \bar{\psi}_{q_a} 
\nn\\
&&
-\psi_{\tilde{q}_a}\,(\Sigma-m_{aR}) \,\psi_{q_a}
-\bar{\psi}_{\tilde{q}_a}\,(\Sigma-m_{aR}) \,\bar{\psi}_{q_a}
\nn \\
&&+i\sqrt{2}\biggl\{
(\psi_{\tilde{q}_a}\psi -\bar{\psi}_{q_a}\bar{\lambda}) 
q_a
-(\psi_{\tilde{q}_a} \lambda 
+\bar{\psi}_{q_a}\bar{\psi}) \tilde{q}^\ast_a
\nn \\
&&
+(\psi_{q_a}\lambda - \bar{\psi}_{\tilde{q}_a}\bar{\psi}) 
q^\ast_a
+(\psi_{q_a}\psi +\bar{\psi}_{\tilde{q}_a}\bar{\lambda}) 
\tilde{q}_a
\biggr\} 
\biggr]
\nn\\
&
=
&
-\frac{1}{2g^2}\bar\lambda_i{\gamma}^M\partial_M{\lambda}^i
+\sum_{a=1}^n\biggl[-\bar{\psi}_a{\gamma}^M\partial_M\psi_a 
-\bar\psi_a(\Sigma-m_{aR})\psi_a
\nn \\
&&-i\sqrt{2}
\bar\psi_{a}\lambda^i \epsilon_{ij}q^j_a 
+i\sqrt{2}
\bar\lambda_i\psi_{a} \epsilon^{ij}q^\ast_{ja} 
\biggr]
\, ,
\label{eq:5D-lag-fermion}
\end{eqnarray}
where capitalized indices $M, N, \dots$ run over $0,1,2,3,5$. 
The gamma matrices in five dimensions are given by $4\times 4$ 
matirces as 
\begin{equation}
\gamma^M=\left(\,\left(\begin{array}{cc}0&\sigma^m\\ \bar{\sigma}^m&0
\end{array}\right),\left(\begin{array}{cc}-i&0\\ 0&i\end{array}\right)\,
\right)\,,
\end{equation}
where $\sigma^m=(1,\vec{\sigma})$ and $\bar{\sigma}^m=(1,-\vec{\sigma})$ 
\cite{WB}. 
To achieve the ${\cal N}=2$ SUSY off-shell formalism, it is 
convenient to make $SU(2)_R$ manifest. 
In the case of vector multiplet, the spinors 
in five dimensions are most conveniently organized 
in terms of the 
symplectic ($SU(2)$) Majorana 
spinors $\lambda^i$,  $i=1,2$, transforming as doublets 
under the $SU(2)_R$ symmetry. 
The $SU(2)$ Majorana spinor is defined by 
\begin{equation}
\lambda^i=\epsilon^{ij}C\bar{\lambda}_j^T\,,
\end{equation}
where the charge conjugation matrix $C$ in five dimensions 
satisfies $C\gamma^MC^{-1}=(\gamma^M)^T$, 
$C^T=-C$, and $C C^\dagger =1$. 
An explicit form may be given by $C=$diag$(i\sigma^2,i\sigma^2)$. 
The spinors in the ${\cal N}=2$ vector multiplet can be assembled 
into a four-component $SU(2)$ Majorana spinor $\lambda^i$ as 
\begin{equation}
\lambda^1
=
\left(\begin{array}{c}
\lambda_\alpha\\ \bar{\psi}^{\dot{\alpha}}
\end{array}\right)\,,\quad
\lambda^2
=\left(\begin{array}{c}
\psi_\alpha\\ -\bar{\lambda}^{\dot{\alpha}}
\end{array}\right)\,,
\end{equation}
\begin{equation}
\bar{\lambda}_1
=\left(\begin{array}{cc}
\psi^\alpha & \bar{\lambda}_{\dot{\alpha}}
\end{array}\right) \,,\quad
\bar{\lambda}_2=\left(\begin{array}{cc}
-\lambda^\alpha & \bar{\psi}_{\dot{\alpha}}
\end{array}\right) \,.
\end{equation}
The spinor in the hypermultiplets are singlets under the 
$SU(2)_R$ symmetry and is assembled into a four-component spinor 
$\psi_a$ for each flavor 
\begin{equation}
\psi=
\left(\begin{array}{c}
\psi_{q a} \\ \bar{\psi}^{\tilde{q}_a}
\end{array}\right)\,,\quad
\bar{\psi}
=\left(\begin{array}{cc}
\psi_{\tilde{q}_a} & -\bar{\psi}_{qa}
\end{array}\right) \,.
\end{equation}
The scalars and auxiliary fields 
in the hypermultiplet transform as doublet 
under the $SU(2)_R$ 
as 
given in Eqs.(\ref{sym4}) and (\ref{sym7}).

The above five-dimensional Lagrangian 
(\ref{5D-lagrangian})--(\ref{eq:5D-lag-fermion}) 
makes it clear that we can have only real mass 
parameters for hyeprmultiplets, 
which is obtained as a momentum 
in one extra dimension 
(the sixth dimension) through a nontrivial 
(Scherk-Schwarz) 
\cite{ScherkSchwarz} dimensional reduction 
from six dimensions \cite{SierraTownsend}. 
On the other hand, we need to have at least three 
discrete vacua in complex field plane. 
This situation can be realized in our model through 
the complex masses of hypermultiplets. 
This is the reason why we cannot generalize our junction 
solution to five or six-dimensions.

%%%%%%%%%%%%%%%%%%%%%%%%%%%%%%%%%%%%%%%%%%%%%%%%%%%%%%%%
%%%%%%%%%%%%%%%%%%%%%%%%%%%%%%%%%%%%%%%%%%%%%%%%%%%%%%%%
\renewcommand{\thesubsection}{Acknowledgments}
\subsection{}

One of the authors (N.S.) acknowledges useful 
discussion of ${\cal N}=2$ SUSY theories 
with Masato Arai, Masashi Naganuma, Muneto Nitta, 
and Keisuke Ohashi. 
This work is supported in part by Grant-in-Aid 
 for Scientific Research from the Japan Ministry 
 of Education, Science and Culture 13640269. 

%================================================
% References
%
%================================================

%%%%%%%%%% References %%%%%%%%%%%%%%%%%%%%%%%%%
\newcommand{\J}[4]{{\sl #1} {\bf #2} (#3) #4}
\newcommand{\andJ}[3]{{\bf #1} (#2) #3}
\newcommand{\AP}{Ann.\ Phys.\ (N.Y.)}
\newcommand{\MPL}{Mod.\ Phys.\ Lett.}
\newcommand{\NP}{Nucl.\ Phys.}
\newcommand{\PL}{Phys.\ Lett.}
\newcommand{\PR}{ Phys.\ Rev.}
\newcommand{\PRL}{Phys.\ Rev.\ Lett.}
\newcommand{\PTP}{Prog.\ Theor.\ Phys.}
\newcommand{\hep}[1]{{\tt hep-th/{#1}}}
%%%%%%%%%%%%%%%%%%%%%%%%%%%%%%%%%%%%%%%%%%%%%%%

\end{document}